\documentclass[aps,prd,letterpaper,showkeys,showpacs,onecolumn,nofootinbib,preprintnumbers,longbibliography,unsortedaddress]{revtex4-1}

\usepackage[top=2.8cm, bottom=2.8cm, left=2cm, right=2cm]{geometry}
\usepackage[utf8]{inputenc}

\usepackage{amsmath,amssymb,lmodern,amsfonts} 
\usepackage{bm,latexsym,amsmath,amssymb,amsfonts,mathrsfs,ulem}
\usepackage[dvipdfmx]{graphicx}
\usepackage{natbib}
\usepackage{hyperref} 
\usepackage{color}
\usepackage{tabulary} 
\usepackage{empheq}
\usepackage[x11names]{xcolor}
\usepackage[nodisplayskipstretch]{setspace}

\newcommand{\be}{\begin{equation}}
\newcommand{\ee}{\end{equation}}
\newcommand{\bea}{\begin{eqnarray}}
\newcommand{\eea}{\end{eqnarray}}

\def\ba{\begin{array}}
\def\ea{\end{array}}
\def\bfig{\begin{figure}}
\def\efig{\end{figure}}

\def\tB{\hat B}

\def\Bn{B_*}

\def\dotBn{\dot B_*}

\newcommand*{\A}{{\cal A}}
\newcommand*{\B}{{\cal B}}
\newcommand*{\C}{{\cal C}}
\newcommand*{\D}{{\cal D}}
\newcommand*{\E}{{\cal E}}
\newcommand*{\F}{{\cal F}}
\newcommand*{\M}{{\cal M}}

\newcommand*{\Bs}{B_*{}}

\newcommand*{\Bh}{{\hat B}}

\newcommand*{\Zs}{Z_*{}}

\newcommand*{\Zh}{{\hat Z}}

\def\beq{\begin{equation}}
\def\eeq{\end{equation}}
\def\bi{\begin{itemize}}
\def\ei{\end{itemize}}
\def\ba{\begin{array}}
\def\ea{\end{array}}
\def\bfig{\begin{figure}}
\def\efig{\end{figure}}

\def\tB{\hat B} 
 
\def\A{{\cal A}}
\def\B{{\cal B}}
\def\C{{\cal C}}

\newcommand{\eqn}[1]{(\ref{#1})}
\newcommand{\eq}[1]{Eq.~(\ref{#1})}

\begin{document}

\preprint{PI/UAN-2021-704FT}

\title{Towards the extended SU(2) Proca theory}

\author{Alexander Gallego Cadavid}
\email{alexander.gallego@uv.cl}
\affiliation{Instituto de F\'{\i}sica y Astronom\'{\i}a,  Universidad  de Valpara\'{\i}so, \\ Avenida Gran Breta\~na 1111,  Valpara\'{\i}so  2360102,  Chile}

\author{Carlos M. Nieto}
\email{caniegue@correo.uis.edu.co}
\affiliation{Escuela  de  F\'isica,  Universidad  Industrial  de  Santander, \\ Ciudad  Universitaria,  Bucaramanga  680002,  Colombia}

\author{Yeinzon Rodr\'iguez}
\email{yeinzon.rodriguez@uan.edu.co}
\affiliation{Centro de Investigaciones en Ciencias B\'asicas y Aplicadas, Universidad Antonio Nari\~no, \\ Cra 3 Este \# 47A-15, Bogot\'a D.C. 110231, Colombia}
\affiliation{Escuela  de  F\'isica,  Universidad  Industrial  de  Santander, \\ Ciudad  Universitaria,  Bucaramanga  680002,  Colombia}

\begin{abstract}
In this work, we explore the construction of the most general vector-tensor theory with an SU(2) global symmetry in the vector sector as a proposal for a modified theory of gravity. We start with a general Lagrangian containing terms involving symmetric and/or antisymmetric combinations of the covariant derivative of the vector field plus an arbitrary function of the vector field times the Ricci scalar. Then, we study the degeneracy of the full theory to determine whether it can be healthy or not. We find relations among some of the free functions in the Lagrangian that are necessary for the healthiness of the theory in correspondence with the several ways in which the kinetic matrix can be turned degenerate. Finally, we take the decoupling limit of the theory and find additional conditions on the free functions that are necessary for the healthiness of the longitudinal modes.
\end{abstract}



\maketitle

\section{Introduction}

It is the common wisdom in science that, at the end, all theories are effective.  In their purpose of describing Nature in the most faithful way, each scientific theory comes with its own built-in range of applicability, and General Relativity (GR) is not the exception. Perhaps, the most emblematic phenomenon in GR that signals the breakdown of the theory is the existence of singularities.  Whether a consistent quantum gravity theory or just a modification of the classical gravitational theory is needed to bypass the failures of GR, it is something unknown at the moment.  Therefore, the quest for advancement in the understanding of the gravitational interaction calls for an exploration of both alternatives. 

In the world of the (classical) modified gravity theories, Lovelock's theorem \cite{Lovelock:1971yv,Lovelock:1972vz} in the early 70's gave the community a strong lesson: there is no way to modify GR in four dimensions if the action is to be constructed with the metric and its first- and second-order derivatives and the field equations are to be of second order.  Thus, one of the most reasonable ways to proceed is to add more gravitational degrees of freedom, for example, a scalar field.  This is the realm of the Horndeski-inspired gravity theories.

The family of Horndeski-inspired theories of modified gravity is growing.  The story began in 1974 when Gregory W. Horndeski constructed the most general scalar-tensor theory that gives rise to second-order field equations \cite{Horndeski:1974wa}.  Two years later, he exported the ideas of his previous work to the case of a vector-tensor theory with a U(1) gauge invariance \cite{Horndeski:1976gi}.  These pioneer works slept well under the dust of time until his scalar-tensor theory was rediscovered in 2009 as the Galileon theory \cite{Nicolis:2008in,Deffayet:2009mn,Deffayet:2011gz,Deffayet:2009wt,Kobayashi:2011nu}.  They being just the next step to extend the Lovelock's results on gravity theories, it is quite surprising that the modified gravity community took almost forty years to pay appropriate attention to Horndeski's efforts.  

Much has been done since 2009 in the study of the theoretical, astrophysical, and cosmological consequences of these theories (see the following interesting reviews and references therein: \cite{Deffayet:2013lga,Rodriguez:2017ckc,Heisenberg:2018vsk,Kobayashi:2019hrl}).  Multi-Galileon theories have been constructed \cite{Hinterbichler:2010xn,Padilla:2010ir,Padilla:2010de,Padilla:2012dx,Sivanesan:2013tba,Kobayashi:2013ina,Ohashi:2015fma,Allys:2016hfl,Akama:2017jsa,Aoki:2021kla}, the $p$-form Galileons have been too \cite{Deffayet:2010zh,Deffayet:2016von,Deffayet:2017eqq}, so has been the generalization to the Proca theory\footnote{This is the extension of the 1976 Horndeski's vector-tensor theory to the case of broken gauge invariance.  The reason for the explicit breaking of the invariance under local U(1) transformations comes from the severe restrictions the latter imposes when trying to implement a vector field as a new gravitational degree of freedom. Horndeski's 1976 construction gives faith of this circumstance. The respective no-go theorem in flat spacetime is presented in Ref. \cite{Deffayet:2013tca}.} \cite{Tasinato:2014eka,Heisenberg:2014rta,Allys:2015sht,Jimenez:2016isa,Allys:2016jaq}, as well as has been the generalization of the SU(2) Yang-Mills theory\footnote{This is also known as the generalized SU(2) Proca theory (GSU2P).} \cite{Allys:2016kbq,GallegoCadavid:2020dho,GallegoCadavid:2022uzn,Jimenez:2016upj} (see also Ref. \cite{Nicosia:2020egv}).  Even more, the generalized scalar-vector-tensor theory that merges the Horndeski theory with the generalized Proca theory (GP) has been built quite recently \cite{Heisenberg:2018acv}.  It was later recognized that the Ostrogradski-ghost-free theories space\footnote{Ostrogradski's ghost is responsible for a Hamiltonian unbounded from below \cite{Ostrogradsky:1850fid}, see also Refs. \cite{Woodard:2006nt,Woodard:2015zca}.} is much wider.  Having second-order field equations for the purely scalar, vector, and tensor sectors of these theories, as well as for the mixed sectors, is desirable, very convenient indeed, but unnecessary to avoid the hateful Ostrogradski's ghost as long as the respective Lagrangians are degenerate.  Thus, several new terms were added to the original Lagrangians, terms that give rise to higher-order field equations without spoiling the healthiness of the theories.  Such new terms were collectively called ``beyond'':  beyond Horndeski terms \cite{Gleyzes:2014dya,Gleyzes:2014qga}, beyond Proca terms \cite{Heisenberg:2016eld,GallegoCadavid:2019zke}, and beyond SU(2) Proca terms \cite{GallegoCadavid:2020dho,GallegoCadavid:2022uzn}.  

The complete set of Lagrangian pieces that preserve the healthiness of the scalar-tensor theory no matter if the field equations are second order or not gives shape to what is called the extended scalar-tensor theory or the degenerate higher-order scalar-tensor theory (DHOST) \cite{Langlois:2015cwa,Achour:2016rkg,BenAchour:2016fzp,Crisostomi:2016czh}.  The degeneracy of the kinetic matrix corresponds to the primary constraint-enforcing relation required for the propagation of the right number of degrees of freedom.  This is not enough, in general, but it is for DHOST \cite{Langlois:2015skt}.  In the similar fashion, the complete set of Lagrangian pieces that preserve the healthiness of the vector-tensor theory was intended to be obtained in Ref. \cite{Kimura:2016rzw}, its authors having reached partial success.  This is because there is no proof yet that the primary constraint-enforcing relation is enough to close the constraint algebra \cite{ErrastiDiez:2019ttn,ErrastiDiez:2019trb,ErrastiDiez:2020dux}.  Moreover, the authors did not investigate whether the mixed and non-mixed sectors of the theory are healthy or not.  Something similar happened with the construction of the GSU2P\footnote{As with the GP, the construction of a gauge-invariant GSU2P is highly restrictive and actually not necessary:  the GSU2P pretends just to be an effective theory without any connection to the gauge theories that describe the electromagnetic and nuclear fundamental interactions. The global invariance of this theory under SU(2) internal transformations has actually to do with the natural realization of the cosmological principle (see Refs. \cite{GallegoCadavid:2020dho,Garnica:2021fuu}). On the other hand, the invariance under global transformations presents quite frequently in Nature;  for instance, Special Relativity, as an effective theory of space and time, is a theory invariant under global Poincaré transformations.}:  in Ref. \cite{Allys:2016kbq}, the action was built having in mind just the primary constraint-enforcing relation which turned out not to be sufficient because a secondary constraint-enforcing relation was unveiled in Refs. \cite{ErrastiDiez:2019ttn,ErrastiDiez:2019trb}.  Several terms were discarded in the original GSU2P because they were equivalent to others in the Lagrangian up to total derivatives;  unfortunately, those total derivatives did not satisfy the secondary constraint-enforcing relation, forcing the reconstruction of the GSU2P from scratch, a task that was carried out successfully in Refs. \cite{GallegoCadavid:2020dho,GallegoCadavid:2022uzn}.  

It is the purpose of this paper to give a step ahead in the construction of the extended version of the GSU2P.  To this end, we follow the techniques of Refs. \cite{Langlois:2015cwa,Kimura:2016rzw} to build the version of the GSU2P that implements the primary constraint-enforcing relation in the most general possible way\footnote{The GP and the GSU2P implement the primary constraint-enforcing relation so that the temporal component of the vector field is the one that does not propagate.  In the extended vector-tensor theory (EVT) \cite{Kimura:2016rzw} as well as in the extended version of the GSU2P we want to start building in this paper, the non-propagating degree of freedom can be any linear combination of components of the vector field, see also Refs. \cite{BeltranJimenez:2019wrd,deRham:2020yet}.}.  To make it less difficult, although it is already difficult enough, we concentrate just in the sector of the theory that involves two first-order derivatives of the vector field with the addition of a non-minimal coupling to the Ricci scalar via an arbitrary function of the vector field.  We impose additional restrictions to the obtained theory coming from the requirement that the decoupling limit of the Lagrangian is degenerate as well.  We expect to complete our task in the future by investigating the addition of more terms to the Lagrangian that exhibit non-minimal couplings to the curvature\footnote{We already know from Ref. \cite{GallegoCadavid:2020dho} that couplings to both the Riemann tensor and the Einstein tensor are non trivial.}, the  constraint algebra of the theory, and the other mixed and non-mixed sectors of it. Interesting astrophysical and cosmological applications surely await us, such as those already explored for the EVT \cite{Kase:2018tsb} and the GSU2P \cite{Rodriguez:2017wkg,Garnica:2021fuu}\footnote{The cosmological inflationary application of the GSU2P \cite{Garnica:2021fuu} is indeed very interesting: constant-roll inflation is realized as an attractor curve whose attraction basin covers most of the available phase space.  This contributes significantly to the solution of the inflationary initial conditions problem.  In addition, the scenario is free of Big-Bang singularities, and can help with the production of primordial black holes that might serve as dark matter as described in Ref. \cite{Motohashi:2019rhu}.}.

The layout of the paper is as follows:  in Section \ref{extsec}, we introduce the extended SU(2) Proca action, split conveniently into four pieces.  The 3+1 decomposition of this action is performed in Section \ref{decomposition}.  In Section \ref{basischange}, a clever change of basis is carried out so that it is much less difficult to obtain the conditions for the degeneracy of the kinetic matrix.  The latter is performed in Section \ref{degeneracycond}.  The same recipe is followed for the decoupling limit of the theory in Section \ref{scalarlim}. Finally, the conclusions are presented in Section \ref{conclusions}.

Throughout the text, Greek indices are space-time indices and run from 0 to 3, while Latin indices, the first of the alphabet, label both internal SU(2) group indices and the 
vectors in the two bases introduced in the text. These Latin indices run from 1 to 3.  The sign convention is the (+++) according to Misner, Thorne, and Wheeler \cite{Misner:1974qy}.

\section{The Extended SU(2) Proca action} \label{extsec}
In this section, we present part of the action of the new Extended SU(2) Proca theory. To some degree, we follow the seminal works in Refs.~\cite{Langlois:2015cwa} and ~\cite{Kimura:2016rzw} (scalar-tensor and vector-tensor degenerate theories, respectively). Nonetheless, we extend the procedure in those papers to the case of non-Abelian vector-tensor theories. 

First, in order to simplify the discussion, we split the action into four pieces, three of them being $S_{A}$, $S_{S}$, and the mixed term $S_{AS}$, corresponding to terms proportional to the antisymmetric tensor $A^{a}_{\mu\nu} \equiv \nabla_{\mu}B^{a}_{\nu}-\nabla_{\nu}B^{a}_{\mu}$, the symmetric tensor $S^{a}_{\mu\nu}  \equiv \nabla_{\mu}B^{a}_{\nu}+\nabla_{\nu}B^{a}_{\mu}$, and the combinations between $A^{a}_{\mu\nu}$ and $S^{a}_{\mu\nu}$, respectively.  In the previous definitions, $\nabla_\mu$ is the space-time covariant derivative, and $B_\mu^a$ is the vector field that belongs to the Lie algebra of the SU(2) group of global transformations under which the action is made invariant.  The fourth piece corresponds to a non-minimal coupling of $B_\mu^a$ to gravity via the Ricci scalar.

We begin with the most general action constructed from two powers of the antisymmetric tensor $A^{a}_{\mu\nu}$ and that is invariant under global SU(2) transformations:
\begin{equation}\label{eq:sa}
 S_{A}=\int d^{4}x\sqrt{-g}\left[ C^{\mu\nu\rho\sigma}_{ab}A^{a}_{\mu\nu}A^{b}_{\rho\sigma}\right] \,,
\end{equation}
where $g$ is the determinant of the space-time metric $g_{\mu\nu}$. In this expression, the tensor $C^{\mu\nu\rho\sigma}_{ab}$ is built up out of $g^{\mu\nu}$, the orientability 4-form $\varepsilon^{\mu\nu\rho\sigma}$ in the space-time
manifold, the vector field $B^{a}_{\mu}$, the metric of the SU(2) Lie group (which is proportional to the Kronecker delta), and the orientability 3-form  $\varepsilon_{abc}$ in the SU(2) manifold. Since the numbers of combinations of tensors that can be included in $C^{\mu\nu\rho\sigma}_{ab}$ is quite large, we restrict ourselves to the case where there is at most a second power of $A^{a}_{\mu \nu}$. We can write the tensor $C^{\mu\nu\rho\sigma}_{ab}$ as follows:
\begin{equation}
 C^{\mu\nu\rho\sigma}_{ab} \equiv E^{\mu\nu\rho\sigma}\delta_{ab}+ H^{\mu\nu\rho\sigma}_{ab} \,,
\end{equation}
where 
\be
E^{\mu\nu\rho\sigma} \equiv \alpha_{1}\ g^{\mu\rho}g^{\nu\sigma}
+\alpha_{2}\ \varepsilon^{\mu\nu\rho\sigma} +\alpha_{3}\ \varepsilon^{\mu\rho\sigma\beta}B^{\nu}_c B^{c}_{\beta}+\alpha_{4}\ g^{\mu\beta}g^{\nu\sigma}g^{\rho\alpha}
 B_{\alpha c}B^{c}_{\beta}
\label{EA} \,,
\ee
and
\bea
H^{\mu\nu\rho\sigma}_{ab}&\equiv&\beta_{1}\ g^{\nu\rho}\varepsilon^{\mu\sigma\alpha\beta}
 B_{\alpha a}B_{\beta b} +\beta_{2}\ \varepsilon^{\mu\rho\sigma\beta}B^{\nu}_a B_{\beta b}+\beta_{3}\ \varepsilon^{\mu\rho\sigma\beta}
 B^{\nu}_b B_{\beta a} +\beta_{4}\ \varepsilon^{\mu\nu\rho\sigma}B_{\alpha a}B^{\alpha}_b+\beta_{5}\ g^{\mu\beta}g^{\nu\sigma}g^{\rho\alpha}B_{\alpha a}B_{\beta b} \notag \\
 &&+\beta_{6}\ g^{\mu\beta}g^{\nu\sigma}g^{\rho\alpha}B_{\alpha b}B_{\beta a} +\beta_{7}\ g^{\mu\sigma}g^{\nu\rho}B_{\alpha a}B^{\alpha}_b \,.
 \label{HA}
\eea
The couplings in Eqs. (\ref{EA}) and (\ref{HA}) are functions of $X \equiv B^{a}_{\mu}B_{a}^{\mu}$, i.e.,  $\alpha_i= \alpha_i(X)$ and $\beta_i=\beta_i(X)$. Nonetheless, we omit the argument of these coupling functions, and similar ones later introduced, throughout the paper.

Similarly, we use the symmetric tensor $S_{\mu \nu}^a$ to build the action
\begin{equation}\label{eq:ss}
 S_{S}=\int d^{4}x\sqrt{-g}\left[\tilde{C}^{\mu\nu\rho\sigma}_{ab}S^{a}_{\mu\nu}S^{b}_{\rho\sigma}\right] \,.
\end{equation}
As before, we split the tensor $\tilde{C}^{\mu\nu\rho\sigma}_{ab}$ into two pieces:
\begin{equation}
 \tilde{C}^{\mu\nu\rho\sigma}_{ab} \equiv \tilde{E}^{\mu\nu\rho\sigma}\delta_{ab}+ \tilde{H}^{\mu\nu\rho\sigma}_{ab} \,,
\end{equation}
where
\be \label{Et}
\tilde{E}^{\mu\nu\rho\sigma}=\gamma_{1}\ g^{\mu\nu}g^{\rho\sigma}
+\gamma_{2}\ g^{\mu\rho}g^{\nu\sigma} +\gamma_{3}\ g^{\mu\nu}B^{\rho}_c B^{\sigma c}+\gamma_{4}\ g^{\mu\rho}B^{\nu}_c B^{\sigma c}  \,,
\ee
and
\be
\tilde{H}^{\mu\nu\rho\sigma}_{ab} \equiv \omega_{1}\ g^{\mu\nu}g^{\rho\sigma}B^{\alpha}_{a}B_{\alpha b} +\omega_{2}\ g^{\mu\nu}B^{\rho}_{a}B^{\sigma}_{b} +\omega_{3}\ g^{\mu\rho}g^{\nu\sigma}B^{\alpha}_{a}B_{\alpha b} +\omega_{4}\ g^{\mu\rho}B^{\nu}_{a}B^{\sigma}_{b} +\omega_{5}\ g^{\mu\rho}B^{\sigma}_{a}B^{\nu}_{b} +\omega_{6}\ g^{\nu\rho} \varepsilon^{\mu\sigma\alpha\beta} B_{\alpha a}B_{\beta b} \,.
\ee

Now, we construct an action containing the tensor product of $A^{a}_{\mu\nu}$ and $S^{a}_{\mu\nu}$ as
\begin{equation}\label{eq:sas}
 S_{AS}=\int d^{4}x\sqrt{-g}\left[\hat{C}^{\mu\nu\rho\sigma}_{ab}A^{a}_{\mu\nu}S^{b}_{\rho\sigma}\right] \,.
\end{equation}
Here, the tensor $\hat{C}^{\mu\nu\rho\sigma}_{ab}$ is written as
\begin{equation}
 \hat{C}^{\mu\nu\rho\sigma}_{ab} \equiv \hat{E}^{\mu\nu\rho\sigma}\delta_{ab}+ \hat{H}^{\mu\nu\rho\sigma}_{ab} \,,
\end{equation}
where
\begin{equation}
\hat{E}^{\mu\nu\rho\sigma} \equiv \chi_{1}\ g^{\mu\rho}B^{\nu}_c B^{\sigma c}+\chi_{2}\ \varepsilon^{\mu\nu\sigma\beta} B^{\rho}_c B^c_{\beta} \,,
\end{equation}
and
\bea
\hat{H}^{\mu\nu\rho\sigma}_{ab}&\equiv&\kappa_{1}\ g^{\mu\rho}B^{\nu}_{a}B^{\sigma}_{b} +\kappa_{2}\ g^{\mu\rho}B^{\sigma}_{a}B^{\nu}_{b}  +\kappa_{3}\ g^{\rho\sigma}B^{\mu}_{a}B^{\nu}_{b} \notag \\
&&+\kappa_{4}\ g^{\nu\rho} \varepsilon^{\mu\sigma\alpha\beta} B_{ \alpha a}B_{\beta b} +\kappa_{5}\ g^{\rho\sigma} \varepsilon^{\mu\nu\alpha\beta} B_{\alpha a}B_{\beta b} +\kappa_{6}\ \varepsilon^{\mu\nu\sigma\beta} B^{\rho}_{a}B_{\beta b} +\kappa_{7}\ \varepsilon^{\mu\nu\sigma\beta} B^{\rho}_{b}B_{\beta a} \,.
\eea

Finally, collecting all the contributions from Eqs. \eqn{eq:sa}, \eqn{eq:ss}, and \eqn{eq:sas}, we have that part of the most general non-Abelian vector-tensor theory containing up to first-order derivatives of the vector field $B_\mu^a$ is given by
\begin{equation}\label{Lesu2pt}
 S=\int d^{4}x\sqrt{-g}\left[f(X) R \right] + S_A+S_S+S_{AS} \,,
\end{equation}
where, for simplicity, we have restricted to two powers of the first-order derivatives of the vector field, up to two powers of the vector field itself before multiplying each derivative term by the respective coupling function, and to a non-minimal coupling of the vector field with the Ricci scalar\footnote{There certainly exist non-minimal couplings to gravity involving either the Riemann tensor or the Einstein tensor (see Ref. \cite{GallegoCadavid:2020dho}).  We expect us to consider them in a future publication.}.  Notice that, for $f=1$, the first term in the previous expression is simply the Einstein-Hilbert action (in suitable units). 

In the following sections, we study the degeneracy of the extended SU(2) Proca theory given by the action in \eq{Lesu2pt}. In order to achieve this, we look for some algebraic relations between the coupling functions, $\alpha_i, \beta_i, \gamma_i, \omega_i, \chi_i$ and $\kappa_i$, such that the kinetic matrix of the theory is degenerate (i.e., it has a zero determinant). Thus, our first step in the next section is to decompose the Lagrangian of the theory using the $3+1$ ADM formalism \cite{3+1book}.

\section{3+1 decomposition} \label{decomposition}

In this section, we rewrite the Lagrangian using the $3+1$ decomposition. This procedure is crucial when we want to understand the evolution of a theory for a quite general form of the metric tensor (for instance, when one wants to study the degeneracy properties of a Lagrangian). Following Refs.~\cite{Langlois:2015cwa, Kimura:2016rzw}, we first split the time derivatives from the spatial ones using a covariant 3+1 decomposition of the spacetime; i.e., we do not introduce a coordinate system.
Instead, we work with tensors that are decomposed into time-like and space-like components. 

We  assume the existence of a space-time foliation with 3-dimensional space-like hypersurfaces $\Sigma_{t}$. Then, we introduce a unit vector $n^\mu$ which is both
normal to the hypersurfaces and timelike (i.e., it satisfies the normalization condition $n_\mu n^\mu=-1$) . This foliation induces a three-dimensional metric $h_{\mu\nu}$ on the spatial hypersurfaces  $\Sigma_{t}$ which is defined by:~\cite{Langlois:2015cwa, Kimura:2016rzw,3+1book}
\begin{equation}
h_{\mu\nu} \equiv g_{\mu\nu} + n_{\mu} n_{\nu} \,.
\end{equation}
The SU(2) Proca field $B_{\mu}^a$ can be decomposed using  this induced metric and the normal vector $n^{\mu}$ as
\begin{equation}\label{adecom}
B_{\mu}^a = - n_{\mu} B_{*}^a  + \hat{B}_{\mu}^a \,,
\end{equation}
where $B_{*}^a \equiv n^{\mu}B_{\mu}^a$ and $\hat{B}_{\mu}^a \equiv h_{\mu}^{\nu}B_{\nu}^a$ are the normal and spatial projections of $B_\mu^a$, respectively.

The time direction vector $t^\mu \equiv \partial/\partial t$, associated with a time coordinate $t$ that labels the slicing of space-like hypersurfaces, can also be decomposed as 
\beq
t^\mu =N n^\mu +N^\mu \,,
\eeq
where  $N$ and $N^\mu$ are the lapse function and the shift vector (which is orthogonal to $n^\mu$). We also define the ``time derivative'' of any  spatial tensor as the spatial projection of its Lie derivative  with respect to $t^\mu$ and which we denote by a dot. Thus, in our case, we have 
 \begin{eqnarray}
 \dotBn^a &\equiv& t^\mu \nabla_\mu \Bn^a \,, \\ 
 \dot\tB_\mu^a &\equiv& h_\mu^\nu{\cal L}_t \tB_\nu^a = h_\mu^\nu\, (t^\rho \nabla_\rho \tB_\nu^a+\tB_\rho^a \nabla_\nu t^\rho) \,.
 \end{eqnarray}
 
The covariant derivative of the normal vector can be decomposed into the extrinsic curvature $K_{\mu\nu}$ and acceleration vector $a_\mu$:
\begin{equation}
\nabla_{\mu} n_{\nu} = - n_{\mu} a_{\nu} + K_{\mu\nu} \,,
\end{equation}
where
\begin{eqnarray}
a^{\mu} &\equiv& n^{\nu}\nabla_{\nu}n^{\mu} \,, \\
K_{\mu\nu} &\equiv& h_{\mu}^{\rho} h_{\nu}^{\sigma}\nabla_{\rho}n_{\sigma} = \frac{1}{2N}\left(\dot h_{\mu \nu}-D_\mu N_\nu -D_\nu N_\mu \right) \,, \nonumber \\
&&
\end{eqnarray}
where $D_\mu$ denotes the 3-dimensional  covariant derivative associated with the spatial metric $h_{\mu \nu}$.

Using the above definitions, the covariant derivative of the vector field reads~\cite{Langlois:2015cwa, Kimura:2016rzw}
\be
\nabla_{\mu}B_{\nu}^a = n_{\mu} n_{\nu} (\dot{B}_{*}^a - a^{\rho}\hat{B}_{\rho}^a )  
+ n_{\mu} (- \dot{\hat{B}}_{\nu}^a + K_{\nu}{}^{\rho} \hat{B}_{\rho}^a + a_{\nu}B_{*}^a  ) 
 +n_{\nu}(K_{\mu}{}^{\rho}\hat{B}_{\rho}^a - D_{\mu}B_{*}^a)  
 +D_{\mu}\hat{B}_{\nu}^a  -K_{\mu\nu}B_{*}^a \,,
\ee
such that the only relevant terms for the kinetic part of the Lagrangian (i.e., the ones with the ``time derivative'') are \beq \label{lkin}
(\nabla_{\mu}B_{\nu}^a)_{\rm kin} = n_{\mu} n_{\nu} \dot{B}_{*}^a
+ n_{\mu} (- \dot{\hat{B}}_{\nu}^a + K_{\nu}{}^{\rho} \hat{B}_{\rho}^a ) 
 +n_{\nu} K_{\mu}{}^{\rho}\hat{B}_{\rho}^a  
-K_{\mu\nu}B_{*}^a \,.
\eeq

Thus, the kinetic part of the Lagrangian in \eq{Lesu2pt}, quadratic in $\nabla_\mu B_\nu^a$, can be written as
\be  \label{lkin2}
  \mathcal{L}_{\rm kin}=\mathcal{A}_{ab}\dot{B}^{a}_{*}\dot{B}^{b}_{*} +2\mathcal{B}^{\sigma}_{ab}\dot{B}^{a}_{*}\dot{\hat{B}}^{b}_{\sigma} +2\mathcal{C}^{\alpha\beta}_{a}\dot{B}^{a}_{*}K_{\alpha\beta} + \mathcal{D}^{\nu\sigma}_{ab}\dot{\hat{B}}^{a}_{\nu}\dot{\hat{B}}^{b}_{\sigma}
  +2\mathcal{E}^{\nu\alpha\beta}_{a}\dot{\hat{B}}^{a}_{\nu}K_{\alpha\beta} 
  +\mathcal{F}^{\alpha\beta\rho\sigma}K_{\alpha\beta}K_{\rho\sigma} \,,
\ee
where the expressions for the coefficients 
are given in Appendix \ref{aA}. 
In these expressions, 
we have also considered the contribution of the Ricci scalar given by:~\cite{Langlois:2015cwa, Kimura:2016rzw,3+1book}
\be \label{Rcontribution}
R = {}^{(3)}R + K_{\mu\nu}K^{\mu\nu}-K^2 -2 \nabla_{\mu}(a^\mu -K n^{\mu} ) \,,
\ee
where ${}^{(3)}R$ is the three-dimensional Ricci scalar constructed with $h_{\mu \nu}$ and $K \equiv \gamma^{\mu\nu}K_{\mu\nu}$.

Using the previous definitions, the kinetic Lagrangian can be written in terms of a matrix (which is called the kinetic matrix) as
\begin{equation}\label{km1}
 \mathcal{L}_{\rm kin}=
  \begin{bmatrix}
\dot{B}^{a}_{*} & \dot{\hat{B}}^{a}_{\nu} & K_{\rho\delta} 
    \end{bmatrix}
    \begin{bmatrix}
\mathcal{A}_{ab} & \mathcal{B}^{\sigma}_{ab} & \mathcal{C}^{\alpha\beta}_{a} \\
\mathcal{B}^{\nu}_{ab} & \mathcal{D}^{\nu\sigma}_{ab} & \mathcal{E}^{\nu\alpha\beta}_{a} \\
\mathcal{C}^{\rho\delta}_{b} & \mathcal{E}^{\sigma\rho\delta}_{b} & \mathcal{F}^{\alpha\beta\rho\delta}
    \end{bmatrix} 
        \begin{bmatrix}
\dot{B}^{b}_{*}  \\
\dot{\hat{B}}^{b}_{\sigma}\\
K_{\alpha\beta} 
    \end{bmatrix} \,.
\end{equation}
It is hard to study the degeneracy properties of the kinetic matrix in its present form. Therefore, in the next section we propose a suitable vector-tensor basis in order to have a matrix whose entries are scalars \cite{Langlois:2015cwa, Kimura:2016rzw}. In fact, we will see that such a basis will lead us to a block-diagonal matrix.

\section{Change of basis} \label{basischange}
In this section, we perform a change of bases on the hypersurfaces $\Sigma_t$ and the internal space of SU(2) in order to determine the degeneracy properties of the kinetic matrix in \eq{km1}. 

For the internal space, we propose a basis with three orthonormal vectors $\{W^{a}_{i} \}$. These vectors are chosen such that $W^{a}_{1}$ coincides with the direction of $B^{a}_{*}$ and the other two normal and orthogonal vectors being arbitrary in the vector subspace orthogonal to $W^{a}_{1}$: 
\begin{equation}
\{W^{a}_{i} \}=\left\{W^{a}_{1}\equiv \frac{B^{a}_{*}}{|B_{*}|}, W^{a}_{2}, W^{a}_{3}\right\} \,.
\label{eq:basisgaugeone}
\end{equation}
In the previous expression, $|B_{*}| \equiv \sqrt{\Bs_{a}\Bs^{a}}$. Thus, the orthonormality relation reads
\be
W^{a}_{i}W_{j}{}_{a}=\delta_{i j} \,.
\ee
Notice that the $i,j,k$ indices label in this case the vectors in the basis.

Similarly, we propose an orthonormal basis with three vectors $\{V^{i}_{\mu} \}$ for the three-dimensional hypersurfaces. These vectors are chosen such that $V^{1}_{\mu}$ lies along the direction of $\hat{B}^{1}_{\mu}$ and the other two normal and orthogonal vectors being arbitrary in the vector subspace orthogonal to $V^{1}_{\mu}$:
\begin{equation}
\{V^{i}_{\mu} \}=\left\{V^{1}_{\mu}\equiv \frac{\hat{B}^{1}_{\mu}}{ \sqrt{ \hat{B}^{1}_{\nu}\hat{B}_{1}^{\nu} } }, V^{2}_{\mu}, V^{3}_{\mu}\right\} \,.
\label{eq:basishyperone}
\end{equation}
These three vectors fulfil the relations
\be
V^{i}_{\mu}V^{\mu j}=\delta^{i j} \quad \mbox{ and } \quad n^{\mu} V_{\mu}^i = 0 \,.
\ee
Therefore, all three vectors $V_\mu^i$ are orthogonal to  $n^\mu$. Notice that the super-index ``1" in the left-hand side of the definition of $V_\mu^1$ corresponds to the labeling ``1" of the three vectors of the basis, while the one on the right-hand side corresponds to $a=1$ of $\Bh^a_\mu$.

\subsection{Vector decomposition}
Using the two bases $\{ W_{i}^{a} \}$ and $\{ V^{i}_{\mu} \}$, we can decompose $\Bn^{a}$ and $\hat{B}^{a}_{\mu}$ as
\begin{equation}    
\Bn^{a}= \tilde{\Bn}^{i}\ W^{a}_{i} \quad \mbox{ and } \quad \hat{B}^{a}_{\mu}= \tilde{\hat{B}}^{k}_{i} W^{a}_{k}V^{i}_{\mu} \,,
\label{eq:twobases}
\end{equation}
where $\tilde{\Bn}^{i}$ and $\tilde{\hat{B}}^{k}_{i}$ are the components of $\Bn^{a}$ and $\hat{B}^{a}_{\mu}$ in the bases $\{ W_{i}^{a} \}$ and $\{ V^{i}_{\mu} \}$. Combining \eq{eq:basisgaugeone} with
the first equation of the previous expression, we get
\begin{equation}
B^{a}_{*} = |B_{*}| W^{a}_{1} = \tilde{\Bn}^{1} W^{a}_1 + \tilde{\Bn}^{2} W^{a}_2 +\tilde{\Bn}^{3} W^{a}_3 \,.
\end{equation}
Thus, we see that $\tilde{\Bn}^{2}=\tilde{\Bn}^{3}=0$. Similarly, using
 \eq{eq:basishyperone}, we observe that
\begin{equation}
    \hat{B}^{1}_{\mu}= \sqrt{ \hat{B}^{1}_{\nu}\hat{B}_{1}^{\nu} } \ V^{1}_{\mu}=
    V^{1}_{\mu}(W^{1}_{k}\tilde{\hat{B}}^{k}_{1})+ V^{2}_{\mu}(W^{1}_{k}\tilde{\hat{B}}^{k}_{2})+ V^{3}_{\mu}(W^{1}_{k}\tilde{\hat{B}}^{k}_{3})\,.
\end{equation}
Then, we have $W^{1}_{k}\tilde{\hat{B}}^{k}_{1}= \sqrt{ \hat{B}^{1}_{\nu}\hat{B}_{1}^{\nu} }$ and $W^{2}_{k}\tilde{\hat{B}}^{k}_{2}= W^{3}_{k}\tilde{\hat{B}}^{k}_{3}=0$. Here, we can make a further assumption and set $\tilde{\hat{B}}^{k}_{2}=\tilde{\hat{B}}^{k}_{3}=0$.

\subsection{Tensor decomposition}
 Finally, using the $\{ V^{i}_{\mu} \}$ vector basis, we can define six independent symmetric matrices ($U^I_{\mu\nu}, \ I=1,...,6$): 
\bea
U^1_{\mu\nu} \equiv V^1_\mu V^1_\nu \,, \quad 
U^2_{\mu\nu} &\equiv& \frac{1}{\sqrt{2}} (\gamma_{\mu\nu} - U^1_{\mu\nu}) \,, \quad \quad \quad
U^3_{\mu\nu} \equiv \frac{1}{\sqrt{2}}(V^2_\mu V^2_\nu - V^3_\mu V^3_\nu) \,, \notag\\
U^4_{\mu\nu} \equiv \frac{1}{\sqrt{2}}(V^2_\mu V^3_\nu + V^2_\nu V^3_\mu) \,,  \quad 
U^5_{\mu\nu} &\equiv& \frac{1}{ \sqrt{2} }(V^2_\mu V^1_\nu + V^2_\nu V^1_\mu) \,, \quad 
U^6_{\mu\nu} \equiv \frac{1}{\sqrt{2 } }(V^3_\mu V^1_\nu + V^3_\nu V^1_\mu) \,,
\eea
which satisfy  the relations\footnote{The symmetrization and antisymmetrization operations are normalized.
}
\be 
U^{I}_{\mu\nu}U^{J}{}^{\mu\nu} = \delta^{IJ} \quad \mbox{ and } \quad \gamma^{\mu}{}_{(\rho}\gamma^{\nu}{}_{\sigma)} = \delta_{IJ} U^{I}{}^{\mu\nu} U^{J}_{\rho\sigma} \,.
\ee
In other words, the basis $\{ U^{I}_{\mu\nu} \}$ spans the space of symmetric tensors on $ \Sigma_t $.
Then, we can decompose $K_{\mu\nu}$
 using this tensor basis as
 \be
K_{\mu\nu} = K_I \, U_{\mu\nu}^I \,.
\ee

Introducing the decomposition of $\dot{\Bn}^{a}$, $\dot{\hat{B}}^{a}_{\mu}$, and $K_{\mu\nu}$ in Eq. (\ref{lkin2}), we obtain the following Lagrangian containing only scalar quantities:
\bea\label{lkinbasis}
\mathcal{L}_{kin}&=&{\cal A}_{1}(\dot{\tilde{\Bn}}^{1})^{2} 
+{\cal A}_{2}(\dot{\tilde{\Bn}}^{2})^{2} 
+{\cal A}_{3}(\dot{\tilde{\Bn}}^{3})^{2}
+2{\cal A}_{4}(\dot{\tilde{\Bn}}^{1})(\dot{\tilde{\Bn}}^{2})
+2{\cal A}_{5}(\dot{\tilde{\Bn}}^{1})(\dot{\tilde{\Bn}}^{3})
+2{\cal A}_{6}(\dot{\tilde{\Bn}}^{2})(\dot{\tilde{\Bn}}^{3})
\notag \\
&&+2{\cal B}_{1}(\dot{\tilde{\Bn}}^{1})(\dot{ \hat{\tilde{B}} }^{1}_{1})+2{\cal B}_{2}(\dot{\tilde{\Bn}}^{1})(\dot{ \hat{\tilde{B}} }^{2}_{1})+2{\cal B}_{3}(\dot{\tilde{\Bn}}^{1})(\dot{ \hat{\tilde{B}} }^{3}_{1}) +2{\cal B}_{4}(\dot{\tilde{\Bn}}^{2})(\dot{ \hat{\tilde{B}} }^{1}_{1})+2{\cal B}_{5}(\dot{\tilde{\Bn}}^{3})(\dot{ \hat{\tilde{B}} }^{1}_{1}) \notag \\
&& +2{\cal B}_{6}(\dot{\tilde{\Bn}}^{2})(\dot{ \hat{\tilde{B}} }^{2}_{1})
+{\cal B}_{6}(\dot{\tilde{\Bn}}^{3})(\dot{ \hat{\tilde{B}} }^{3}_{1})\notag \\
&& +2\C_1 K_1(\dot{\tilde{\Bn}}^{1})+2\C_2 K_2(\dot{\tilde{\Bn}}^{1})
+2\C_3 K_1(\dot{\tilde{\Bn}}^{2})+2\C_4 K_2(\dot{\tilde{\Bn}}^{2})
+2\C_5 K_1(\dot{\tilde{\Bn}}^{3})+2\C_6 K_2(\dot{\tilde{\Bn}}^{3})\notag \\
&&
+ \D_1 (\dot{ \hat{\tilde{B}} }^{1}_{1})^{2} 
+ \D_2 (\dot{ \hat{\tilde{B}} }^{2}_{1})^{2}
+ \D_3 (\dot{ \hat{\tilde{B}} }^{3}_{1})^{2}
+ \D_4 (\dot{ \hat{\tilde{B}} }^{1}_{1})(\dot{ \hat{\tilde{B}} }^{2}_{1})
+ \D_5 (\dot{ \hat{\tilde{B}} }^{1}_{1})(\dot{ \hat{\tilde{B}} }^{3}_{1})
+ \D_6 (\dot{ \hat{\tilde{B}} }^{2}_{1})(\dot{ \hat{\tilde{B}} }^{3}_{1}) \notag
\\
&& + \D_7 (\dot{ \hat{\tilde{B}} }^{1}_{2})^{2} 
+ \D_8 (\dot{ \hat{\tilde{B}} }^{2}_{2})^{2}
+ \D_9 (\dot{ \hat{\tilde{B}} }^{3}_{2})^{2}+ \D_{10} (\dot{ \hat{\tilde{B}} }^{1}_{3})^{2} 
+ \D_{11} (\dot{ \hat{\tilde{B}} }^{2}_{3})^{2}
+ \D_{12} (\dot{ \hat{\tilde{B}} }^{3}_{3})^{2} 
 \notag \\
&& + \D_{13} (\dot{ \hat{\tilde{B}} }^{1}_{2})(\dot{ \hat{\tilde{B}} }^{2}_{2})
+ \D_{14} (\dot{ \hat{\tilde{B}} }^{1}_{2})(\dot{ \hat{\tilde{B}} }^{3}_{2})
+ \D_{15} (\dot{ \hat{\tilde{B}} }^{1}_{2})(\dot{ \hat{\tilde{B}} }^{2}_{3})
+ \D_{16} (\dot{ \hat{\tilde{B}} }^{1}_{2})(\dot{ \hat{\tilde{B}} }^{3}_{3})
+ \D_{17} (\dot{ \hat{\tilde{B}} }^{2}_{2})(\dot{ \hat{\tilde{B}} }^{3}_{2})\notag \\
&&+ \D_{18} (\dot{ \hat{\tilde{B}} }^{2}_{2})(\dot{ \hat{\tilde{B}} }^{1}_{3}) 
+ \D_{19} (\dot{ \hat{\tilde{B}} }^{3}_{2})(\dot{ \hat{\tilde{B}} }^{1}_{3})
+ \D_{20} (\dot{ \hat{\tilde{B}} }^{1}_{3})(\dot{ \hat{\tilde{B}} }^{2}_{3})
+ \D_{21} (\dot{ \hat{\tilde{B}} }^{1}_{3})(\dot{ \hat{\tilde{B}} }^{3}_{3})
+ \D_{22} (\dot{ \hat{\tilde{B}} }^{2}_{3})(\dot{ \hat{\tilde{B}} }^{3}_{3}) \notag \\
&&
+ 2\E_1 K_1(\dot{ \hat{\tilde{B}} }^{1}_{1}) +2\E_2 K_2(\dot{ \hat{\tilde{B}} }^{1}_{1}) + 2\E_3 K_1(\dot{ \hat{\tilde{B}} }^{2}_{1}) +2\E_4 K_2(\dot{ \hat{\tilde{B}} }^{2}_{1})
+ 2\E_5 K_1(\dot{ \hat{\tilde{B}} }^{3}_{1}) +2\E_6 K_2(\dot{ \hat{\tilde{B}} }^{3}_{1})\notag\\
&& +2\E_7 K_5(\dot{ \hat{\tilde{B}} }^{1}_{2}) +2\E_8 K_6(\dot{ \hat{\tilde{B}} }^{1}_{2}) 
+2\E_9 K_5(\dot{ \hat{\tilde{B}} }^{2}_{2}) +2\E_{10} K_6(\dot{ \hat{\tilde{B}} }^{2}_{2})
+2\E_{11} K_1(\dot{ \hat{\tilde{B}} }^{3}_{2}) +2\E_{12} K_2(\dot{ \hat{\tilde{B}} }^{3}_{2})\notag \\
&& +2\E_{13} K_5(\dot{ \hat{\tilde{B}} }^{1}_{3}) +2\E_{14} K_6(\dot{ \hat{\tilde{B}} }^{1}_{3}) 
+2\E_{15} K_5(\dot{ \hat{\tilde{B}} }^{2}_{3}) +2\E_{16} K_6(\dot{ \hat{\tilde{B}} }^{2}_{3})
+2\E_{17} K_5(\dot{ \hat{\tilde{B}} }^{3}_{3}) +2\E_{18} K_6(\dot{ \hat{\tilde{B}} }^{3}_{3})\notag \\
&&+ \F_1K_1^2 +\F_2 K_2^2 +2\F_3 K_1K_2 +\F_5(K_3^2+K_4^2)+\F_4(K_5^2+K_6^2) \,,
\eea
where the coefficients are given in Appendix \ref{aB}. We can rewrite this kinetic Lagrangian using two $8 \times 8$ matrices and one $2 \times 2$ matrix as follows:
\bea
\mathcal{L}_{\rm kin}=
\left(
\begin{array}{ccc}
	{\bm m}^T & {\bm m}_1^T & {\bm m}_2^T \\
\end{array}
\right)
\left(
\begin{array}{ccc}
	{\cal M} & 0 & 0 \\
	0 & {\cal M}_1 & 0 \\
	0 & 0 & {\cal M}_2 \\
\end{array}
\right)
\left(
\begin{array}{ccc}
	{\bm m} \\
	{\bm m}_1  \\
	{\bm m}_2  \\
\end{array}
\right)
,
\label{Lkin}
\eea
where we have introduced the vectors\footnote{It is worth recalling that we are using the notation introduced in Eq. (\ref{eq:twobases}).} ${\bm m}\equiv \{\dot{\tilde{\Bs}}^{1},\dot{\tilde{\Bs}}^{2},\dot{\tilde{\Bs}}^{3}, \dot{\hat{\tilde{B}}}_{1}^{\ 1}, \dot{\hat{\tilde{B}}}_{1}^{\ 2},\dot{\hat{\tilde{B}}}_{1}^{\ 3}, K_1, K_2\}$, \\
${\bm m}_1\equiv \{\dot{\hat{\tilde{B}}}_{2}^{\ 1}, \dot{\hat{\tilde{B}}}_{2}^{\ 2}, \dot{\hat{\tilde{B}}}_{2}^{\ 3}, \dot{\hat{\tilde{B}}}_{3}^{\ 1}, \dot{\hat{\tilde{B}}}_{3}^{\ 2}, \dot{\hat{\tilde{B}}}_{3}^{\ 3}, K_5, K_6\}$, and ${\bm m}_2\equiv \{K_3, K_4\}$,
and the matrices $\M$, $\M_1$, and $\M_2$ have the following structures:
\bea
&&{\cal M}\equiv \left(
\begin{array}{cccccccc}
\A_{1} & \A_{4} & \A_{5} & \B_{1} & \B_{2} & \B_{3} & \C_{1} & \C_{2} \\
\A_{4} & \A_{2} & \A_{6} & \B_{4} & \B_{6} & 0 & \C_{3} & \C_{4} \\
\A_{5} & \A_{6} & \A_{3} & \B_{5} & 0 & \B_{6} & \C_{5} & \C_{6} \\
\B_{1} & \B_{4} & \B_{5} & \D_{1} & \D_{4} & \D_{5} & \E_{1} & \E_{2} \\
\B_{2} & \B_{6} & 0 & \D_{4} & \D_{2} & \D_{6} & \E_{3} & \E_{4} \\
\B_{3} & 0 & \B_{6} & \D_{5} & \D_{6} & \D_{3} & \E_{5} & \E_{6} \\
\C_{1} & \C_{3} & \C_{5} & \E_{1} & \E_{3} & \E_{5} & \F_{1} & \F_{3} \\
\C_{2} & \C_{4} & \C_{6} & \E_{2} & \E_{4} & \E_{6} & \F_{3} & \F_{2} \\
\end{array}
\right) \,, \quad 
{\cal M}_1 \equiv \left(
\begin{array}{cccccccc}
	\D_{7} & \D_{13} & \D_{14} & 0 & \D_{15} & \D_{16} & \E_{7} & \E_{8} \\
	\D_{13} & \D_{8} & \D_{17} & \D_{18} & 0 & 0 & \E_{9} & \E_{10} \\
	\D_{14} & \D_{17} & \D_{9} & \D_{19} & 0 & 0 & \E_{11} & \E_{12} \\
	0 & \D_{18} & \D_{19} & \D_{10} & \D_{20} & \D_{21} & \E_{13} & \E_{14} \\
	\D_{15} & 0 & 0 & \D_{20} & \D_{11} & \D_{22} & \E_{15} & \E_{16} \\
	\D_{16} & 0 & 0 & \D_{21} & \D_{22} & \D_{12} & \E_{17} & \E_{18} \\
	\E_{7} & \E_{9} & \E_{11} & \E_{13} & \E_{15} & \E_{17} & \F_{4} & 0 \\
	\E_{8} & \E_{10} & \E_{12} & \E_{14} & \E_{16} & \E_{18} & 0 & \F_{4} \\
\end{array}
\right) \,,
\label{kineticM}
\eea
\bea
\quad {\cal M}_2\equiv \left(
\begin{array}{cc}
	\F_5 & 0 \\
	0 & \F_5 \\
\end{array}
\right) \,.
\label{kineticM2}
\eea
These matrices correspond to the scalar, vector, and tensor sectors of the kinetic matrix respectively.

\section{Degeneracy  of the scalar, vector, and tensor sectors} \label{degeneracycond}
In this section, we investigate the conditions under which the full kinetic matrix is degenerate. Given the structure of this matrix, it being block-diagonal, we see that there are three possible ways of setting the determinant equal to zero (the determinant of each submatrix equal to zero independently). In this paper, we focus on the scalar and the vector sectors. 
Regarding $\mathcal{M}_{2}$, and according to its structure (see Eq. (\ref{kineticM2})), the degeneracy of the tensor sector would lead to a condition which
is basis dependent;  hence, we do not consider the tensor sector in this paper.

\begin{enumerate}
{\it \item Degeneracy of the scalar sector.}

We start with the degeneracy of ${\cal M}$.  In this case, a sufficient condition for its degeneracy is given by
\be
\gamma_1 + \gamma_2 =0 \,, \quad \gamma_3 + \gamma_4 =0 \,, \quad \mbox{and} \quad
\chi_1 = 0 \,. \label{fc}
\ee
Other conditions are possible, of course, but they are quite difficult to obtain because of the high dimension of ${\cal M}$ and the complexity of the matrix entries presented in Appendix \ref{aB}.

The constraints in Eq. (\ref{fc}) produce one zero eigenvalue in ${\cal M}$ which corresponds to a second-class primary constraint in the language of Dirac's algorithm \cite{Dirac:1950pj} (see an interesting description of this algorithm applied to Maxwell and Proca vector fields in Ref. \cite{ErrastiDiez:2019trb}).  This relation removes half a degree of freedom in the scalar sector.

{\it \item Degeneracy of the vector sector.}

A sufficient condition for the degeneracy of $\mathcal{M}_1$ is given by
\be
\alpha_4 =0 \,, \quad 
\gamma_2 + \alpha_1 =0 \,,
\quad \mbox{and} \quad \chi_1 - \gamma_4= 0 \,. \label{sc}
\ee 
As in the previous case, other solutions are possible but are quite difficult to obtain because of the same reasons.

The constraints in Eq. (\ref{sc}) produce two zero eigenvalues which correspond to two second-class primary constraints that remove one degree of freedom in the vector sector.

\end{enumerate}

Overall, the sufficient conditions in Eqs. (\ref{fc}) and (\ref{sc}) remove one and a half degrees of freedom.  The imposition of second-class secondary constraints will remove another one and a half degrees of freedom leading to a total of three degrees of freedom, one in the scalar sector and the other two in the vector sector.  Hence, no comparison is possible between the theory developed here and the GSU2P \cite{GallegoCadavid:2020dho} as the latter was constructed so that the unphysical three degrees of freedom are all removed from the scalar sector. Obtaining the second-class secondary constraints is a non-trivial task as can be deduced from Ref. \cite{ErrastiDiez:2019trb}, and its application to the specific theory is not either (see, for instance, Ref. \cite{GallegoCadavid:2020dho} for the case of the GSU2P).  As such, this is beyond the scope of this paper.

Hence, a sufficient condition to remove three degrees of freedom is given by the combination of the constraints in Eqs. (\ref{fc}) and (\ref{sc}), i.e.,
\begin{equation} \label{eqf}
\alpha_1 = \gamma_1 = - \gamma_2 \quad \mbox{and} \quad \alpha_4 = \gamma_3 = \gamma_4 = \chi_1 = 0 \,. 
\end{equation}

\section{Decoupling limit of the theory} \label{scalarlim}
    In this section, we study the decoupling limit of the constructed theory. Our goal is to check whether the resulting theory is also degenerate. If the decoupling limit is not degenerate, the longitudinal modes of the vector fields considered in the original theory of Eq. (\ref{Lesu2pt}) are not healthy. The action of the decoupling limit is found by using the replacement $B^{a}_{\mu}\rightarrow\nabla_{\mu}\phi^{a}$. Given the structure of $A^{a}_{\mu\nu}$, we realize that both $S_{A}$ and $S_{AS}$ vanish in the decoupling limit. Thus, we find that the action associated to the decoupling limit comes from the first and the third terms in Eq. (\ref{Lesu2pt}):
\begin{align}
 S_{d.l.}=\int d^{4}x\sqrt{-g}\ & 4\left[\phantom{\frac{1}{2}}  \frac{1}{4} f(\nabla^{\mu}\phi^a \nabla_{\mu}\phi_a ) \, R + \gamma_{1}\nabla^{\mu}\nabla_{\mu}\phi_a\nabla^{\rho}\nabla_{\rho}\phi^{a}+\gamma_{2}\nabla_{\mu}\nabla_{\nu}\phi_a\nabla^{\mu}\nabla^{\nu}\phi^{a} \right.
  \notag \\
 & +\gamma_{3}(\nabla^{\rho}\phi_{c}\nabla^{\sigma}\phi^{c})\nabla^{\mu}\nabla_{\mu}\phi_{a}\nabla_{\rho}\nabla_{\sigma}\phi^{a} +\gamma_{4}(\nabla^{\nu}\phi_{c}\nabla^{\sigma}\phi^{c})\nabla_{\mu}\nabla_{\nu}\phi_{a}\nabla^{\mu}\nabla_{\sigma}\phi^{a} \notag \\
 & +\omega_{1}(\nabla^{\alpha}\phi_{a}\nabla_{\alpha}\phi_{b})\nabla^{\mu}\nabla_{\mu}\phi^{a}\nabla^{\rho}\nabla_{\rho}\phi^{b} +\omega_{2}(\nabla^{\rho}\phi_{a}\nabla^{\sigma}\phi_{b})\nabla^{\mu}\nabla_{\mu}\phi^{a}\nabla_{\rho}\nabla_{\sigma}\phi^{b}
\notag \\
 &  +\omega_{3}(\nabla^{\alpha}\phi_{a}\nabla_{\alpha}\phi_{b})\nabla_{\mu}\nabla_{\nu}\phi^{a}\nabla^{\mu}\nabla^{\nu}\phi^{b} +\omega_{4}(\nabla^{\nu}\phi_{a}\nabla^{\sigma}\phi_{b})\nabla_{\mu}\nabla_{\nu}\phi^{a}\nabla^{\mu}\nabla_{\sigma}\phi^{b}
  \notag \\
 &\left. +\omega_{5}(\nabla^{\sigma}\phi_{a}\nabla^{\nu}\phi_{b})\nabla_{\mu}\nabla_{\nu}\phi^{a}\nabla^{\mu}\nabla_{\sigma}\phi^{b}+\omega_{6}(\nabla_{\alpha}\phi_{a}\nabla_{\beta}\phi_{b})\varepsilon^{\mu\sigma\alpha\beta}\nabla_{\mu}\nabla_{\nu}\phi^{a}\nabla^{\nu}\nabla_{\sigma}\phi^{b} \phantom{\frac{1}{2}} \right].	
 \label{eq:slaction}
\end{align}
We see that this action contains second-order derivatives of $\phi^{a}$. Therefore, it corresponds to a scalar multiple-field higher-order theory. In order to study the degeneracy properties of this action, we introduce an auxiliary field which leads to an action with first derivatives only (as is done in Ref. \cite{Langlois:2015cwa}). Therefore, we write $Z^{a}_{\mu}=\nabla_{\mu}\phi^{a}$ in order not to confuse the auxiliary field with the original field $B^{a}_{\mu}$. By doing so, we obtain the action
\begin{align}
 S_{d.l.}=\int d^{4}x\sqrt{-g}\ & 4\left[\phantom{\frac{1}{2}} \frac{1}{4} f(Z^{\mu}_a Z_{\mu}^a) \, R + \gamma_{1}\nabla^{\mu}Z_{\mu a}\nabla^{\rho}Z_{\rho}^{a}+\gamma_{2}\nabla_{\mu}Z_{\nu a}\nabla^{\mu}Z^{\nu a} \right.
\notag \\
 & +\gamma_{3}(Z^{\rho}_c Z^{\sigma c})\nabla^{\mu}Z_{\mu a}\nabla_{\rho}Z_{\sigma}^{a}
 +\gamma_{4}(Z^{\nu}_c Z^{\sigma c})\nabla_{\mu}Z_{\nu a}\nabla^{\mu}Z^{a}_{\sigma} \notag \\
 & +\omega_{1}(Z^{\alpha}_{a}Z_{\alpha b})\nabla^{\mu}Z_{\mu}^{a}\nabla^{\rho}Z_{\rho}^{b} +\omega_{2}(Z^{\rho}_{a}Z^{\sigma}_{b})\nabla^{\mu}Z_{\mu}^{a}\nabla_{\rho}Z_{\sigma}^{b}
 +\omega_{3}(Z^{\alpha}_{a}Z_{\alpha b})\nabla_{\mu}Z_{\nu}^{a}\nabla^{\mu}Z^{\nu b} \notag \\
 &\left. +\omega_{4}(Z^{\nu}_{a}Z^{\sigma}_{b})\nabla_{\mu}Z_{\nu}^{a}\nabla^{\mu}Z^{b}_{\sigma}+\omega_{5}(Z^{\sigma}_{a}Z^{\nu}_{b})\nabla_{\mu}Z_{\nu}^{a}\nabla^{\mu}Z_{\sigma}^{b} \phantom{\frac{1}{2}}
 +\omega_{6}(Z_{\alpha a}Z_{\beta b})\varepsilon^{\mu\sigma\alpha\beta}\nabla_{\mu}Z_{\nu}^{a}\nabla^{\nu}Z_{\sigma}^{b} \phantom{\frac{1}{2}}\right] \,.	
\end{align}
As we did before, we use the $3+1$ formalism to rewrite the covariant derivative of the auxiliary field. Then, we first express the vector field $Z_\mu^a$ using its parallel and orthogonal components with respect to the hypersurfaces: $Z^{a}_{\mu}=-Z^{a}_{*}n_{\mu}+\hat{Z}^{a}_{\mu}$. Following what we did in the vector-tensor theory, we write the covariant derivative of $Z^{a}_{\mu}$ using the ``temporal'' and the ``spatial'' components. Since $\nabla_{\mu}Z^{a}_{\nu}=\nabla_{\nu}Z^{a}_{\mu}$, the temporal derivative of $\hat{Z}^{a}_{\mu}$ can be written in terms of quantities associated to the foliation itself;  therefore, it can be removed from the covariant derivative. Thus, the kinetic Lagrangian can be written in terms of $\dot{Z}^{a}_{*}$ and $K_{\mu\nu}$ only:
\begin{equation}
   \mathcal{L}_{kin}=\tilde{\mathcal{A}}_{ab}\dot{Z}^{a}_{*}\dot{Z}^{b}_{*}
  +\tilde{\mathcal{F}}^{\alpha\beta\rho\sigma}K_{\alpha\beta}K_{\rho\sigma} +2\tilde{\mathcal{C}}^{\alpha\beta}_{a}\dot{Z}^{a}_{*}K_{\alpha\beta} \,.
\end{equation}
We can rewrite this action in a matrix form as follows:
\begin{equation}
 \mathcal{L}_{kin}=
  \begin{bmatrix}
\dot{Z}^{a}_{*} & K_{\rho\sigma} 
    \end{bmatrix}
    \begin{bmatrix}
\tilde{\mathcal{A}}_{ab} & \tilde{\mathcal{C}}^{\alpha\beta}_{a} \\	
\tilde{\mathcal{C}}^{\rho\sigma}_{b} & \tilde{\mathcal{F}}^{\alpha\beta\rho\sigma}
    \end{bmatrix} 
        \begin{bmatrix}
\dot{Z}^{b}_{*}  \\
K_{\alpha\beta} 
    \end{bmatrix},
    \label{eq:kinm}
\end{equation}
where the entries of the kinetic matrix are given in Appendix \ref{aC}.

Following the methodology outlined for the vector-tensor theory, we introduce two vector bases:  one on the hypersurface and one in the internal SU(2) space. For the internal space, we have the set of vectors
\begin{equation}
\{W^{a}_{j}\}=\left\{W^{a}_{1}\equiv \frac{Z^{a}_{*}}{|Z_{*}|}, W^{a}_{2}, W^{a}_{3}\right\} \ \mbox{such that} \ W^{a}_{k}W^{l}_{a}=\delta^{l}_{k} \,.
\label{eq:basisgauge}
\end{equation}
On the hypersurfaces, we have
\begin{equation}
\{V^{j}_{\mu}\}=\left\{V^{1}_{\mu}\equiv \frac{\hat{Z}^{1}_{\mu}}{ \sqrt{\hat{Z}^{1}_{\mu} \hat{Z}^\mu_1}},V^{2}_{\mu}, V^{3}_{\mu}\right\} \ \mbox{such that} \ V^{i}_{\mu}V^{\mu}_{j}=\delta^{i}_{j} \,.
\label{eq:basishyper}
\end{equation}
Using the two bases in Eqs. (\ref{eq:basisgauge}) and (\ref{eq:basishyper}), we write the components of the vector field as follows:
\begin{equation}
    Z_{*}^{a}= \tilde{Z_{*}}^{i} W^{a}_{i} \,, \quad \hat{Z}^{a}_{\mu}= \tilde{\hat{Z}}^{k}_{i} W^{a}_{k}V^{i}_{\mu} \,.
\end{equation}
According to the choice of basis vectors, Eqs. (\ref{eq:basisgauge}) and (\ref{eq:basishyper}), we have
\begin{equation}
Z^{a}_{*} = |Z_{*}| W^{a}_{1} = \tilde{Z_{*}}^{1} W^{a}_1 + \tilde{Z_{*}}^{2} W^{a}_2 +\tilde{Z_{*}}^{3} W^{a}_3 \,,
\end{equation}
and
\begin{equation}
    \hat{Z}^{1}_{\mu}= \sqrt{\hat{Z}^{1}_{\mu} \hat{Z}^\mu_1} \, V^{1}_{\mu}=
    V^{1}_{\mu}(W^{1}_{k}\tilde{\hat{Z}}^{k}_{1})+ V^{2}_{\mu}(W^{1}_{k}\tilde{\hat{Z}}^{k}_{2})+ V^{3}_{\mu}(W^{1}_{k}\tilde{\hat{Z}}^{k}_{3}) \,. 
\end{equation}
Without loss of generality, we can make the following assumptions:
$\tilde{Z_{*}}^{2}=\tilde{Z_{*}}^{3}=0$, $W^{1}_{k}\tilde{\hat{Z}}^{k}_{2}=0$, and
$W^{1}_{k}\tilde{\hat{Z}}^{k}_{3}=0$, where, from the last two expressions, we can make $\tilde{\hat{Z}}^{k}_{2}=\tilde{\hat{Z}}^{k}_{3}=0$.
Thus, after employing these assumptions, we write the 
kinetic Lagrangian as
\bea
\mathcal{L}_{\rm kin}=
\left(
\begin{array}{cc}
	\tilde{{\bm u}}_{1}^T & \tilde{{\bm u}}_{2}^T \\
\end{array}
\right)
\left(
\begin{array}{cc}
	\tilde{\mathcal{M}}_{1} & 0  \\
	0 & \tilde{\mathcal{M}}_{2}  \\
\end{array}
\right)
\left(
\begin{array}{cc}
	\tilde{{\bm u}}_{1} \\
	{\bm \tilde{{\bm u}}_{2}}  \\
\end{array}
\right)
,
\label{Lkins}
\eea
where the vector components are
$\tilde{{\bm u}}_{1}\equiv(\dot{\tilde{Z}}^{1}_{*},\dot{\tilde{Z}}^{2}_{*},\dot{\tilde{Z}}^{3}_{*},K_{1},K_{2})$ 
and $\tilde{{\bm u}}_{2}\equiv(K_{5},K_{6}, K_{3},K_{4})$. On the other hand, each entry of the block-diagonal matrix in Eq. (\ref{Lkins}) has the structure
\begin{equation}\label{M1tilde}
\tilde{\M}_{1}=
    \begin{bmatrix}
\tilde{\A}_{1}  & \tilde{\A}_{2} & \tilde{\A}_{3} & \tilde{\C}_{1} & \tilde{\C}_{2} \\
\tilde{\A}_{2} & \tilde{\A}_{4} & \tilde{\A}_{5} & \tilde{\C}_{3} & \tilde{\C}_{4} \\
\tilde{\A}_{3} & \tilde{\A}_{5} & \tilde{\A}_{6} & \tilde{\C}_{5} & \tilde{\C}_{6} \\
\tilde{\C}_{1} & \tilde{\C}_{3} & \tilde{\C}_{5} & \tilde{\F}_{1} & \tilde{\F}_{2} \\
\tilde{\C}_{2} & \tilde{\C}_{4} & \tilde{\C}_{6} & \tilde{\F}_{2} & \tilde{\F}_{3}
    \end{bmatrix},
\end{equation}

\begin{equation}\label{M2tilde}
\tilde{\M}_{2}=
    \begin{bmatrix}
\tilde{\F}_{4} & 0 & 0 & 0\\
0 & \tilde{\F}_{4}  & 0 & 0\\
0 & 0 & \tilde{\F}_{5} & 0 \\
0 & 0 & 0 & \tilde{\F}_{5} 
    \end{bmatrix},
\end{equation}
where the entries of $\tilde{\mathcal{M}}_{1}$ and $\tilde{\mathcal{M}}_{2}$ are given in Appendix \ref{aD}.

Since the kinetic matrix in Eq. (\ref{Lkins}) is a block-diagonal matrix, we have two options for setting the full matrix degenerate. 

A sufficient condition to get $\det \tilde{\mathcal{M}}_{1}=0$ is given by
\begin{equation}
 \gamma_{1} + \gamma_{2} =0 \quad \mbox{ and } \quad \gamma_{3} + \gamma_{4} =0 \,,
 \label{eq:cone}
\end{equation}
which corresponds to the vanishing of just one eigenvalue. There exist more possibilities to make $\det \tilde{\mathcal{M}}_{1}=0$, but the dimension of the matrix and the complexity of the expressions in Appendix \ref{aD} allowed us to find out just this condition. 

On the other hand, if we impose $\det \tilde{\mathcal{M}}_{2}=0$, we obtain the necessary and sufficient condition
\begin{equation}
 \omega_{3} =0 \,, \quad \gamma_2 = \frac{1}{4 X} f \,, \quad \mbox{ and } \quad \gamma_{4} = \omega_{5} \,, \label{eq:cone2}
\end{equation}
where, in this case, $X\equiv Z^{\mu}_a Z_{\mu}^a$ since we are considering the decoupling limit and we have used $X =  -Z_{*}^{2} + \hat{Z}^2$. This condition corresponds to the vanishing of two eigenvalues.

Since there exist three longitudinal modes in the decoupling limit that experience higher-order field equations, it is necessary to remove three ghost degrees of freedom.  The conditions in Eqs. (\ref{eq:cone}) and (\ref{eq:cone2}) correspond to second-class primary constraints that, therefore, remove one and a half degrees of freedom. The conservation in time of these primary constraints will lead to other three, secondary and second-class, constraints that remove another one and a half degrees of freedom. However, the respective Hamiltonian analysis and the obtaining and application of the secondary constraint-enforcing relation are beyond the scope of this paper.

\section{Conclusions} \label{conclusions}

What we have done in this paper is to give a step forward in our intention of building the extended SU(2) vector-tensor theory that implements all the requirements to be considered a healthy theory.  We have written all the possible simultaneously diffeomorphism- and group-invariant terms that can be written with a product of two first-order derivatives of the vector field $B_\mu^a$ and a coupling function of $X\equiv B_\mu^a B^\mu_a$, and we have added them up to the non-minimal coupling of an arbitrary funcion of $X$ to the Ricci scalar.  We have found out the conditions for the degeneracy of the kinetic matrix that guarantee the implementation of the primary constraint-enforcing relation \cite{Langlois:2015cwa,Kimura:2016rzw} which, in turn, is a necessary condition for the propagation of the right number of degrees of freedom.  We have also checked under which conditions the decoupling limit of this theory also has a degenerate kinetic matrix.  Bringing together Eqs. (\ref{fc}), (\ref{sc}), (\ref{eq:cone}), and (\ref{eq:cone2}), we conclude there exist at least one class of theories that satisfy our requirements and that is given by the combination of Eqs. (\ref{eqf}), (\ref{eq:cone}), and (\ref{eq:cone2}):
\begin{equation}
\alpha_1 = \gamma_1 = - \gamma_2 = -\frac{1}{4X} f \,, \quad \mbox{and} \quad \alpha_4 = \gamma_3 = \gamma_4 = \chi_1 = \omega_3 = \omega_5 = 0 \,.    
\end{equation}

It is clear that this is not the end of the story.  The constraint algebra must be studied for this class of theories and its respective decoupling limit.  Only that way, as was done for the DHOST in Ref. \cite{Langlois:2015skt} and for the complete Maxwell-Proca theory in Refs. \cite{ErrastiDiez:2019ttn,ErrastiDiez:2019trb}, we could be sure the constructed theories do propagate the right number of degrees of freedom and, therefore, are free of the Ostrogradski's instability.  We expect us to address this issue in a future publication.  We also expect us to address the rich astrophysical and cosmological consequences of this theory and compare them with those of the EVT \cite{Kase:2018tsb} and the GSU2P \cite{Garnica:2021fuu}.

\section*{Acknowledgments} 
 We acknowledge the anonymous referee for their crucial questioning that made us improve this paper in a fundamental way. A. G. C. was funded by Agencia Nacional de Investigación y Desarrollo  ANID through the FONDECYT postdoctoral Grant No. 3210512. C. M. N. was supported by Vicerrectoría de Investigación y Extensión – Universidad Industrial de Santander
Postdoctoral Fellowship Programme No. 2021000126. Y. R. has received funding/support from the Patrimonio Autónomo - Fondo Nacional de Financiamiento para la Ciencia, la Tecnología y la Innovación Francisco José de Caldas (MINCIENCIAS - COLOMBIA) Grant No. 110685269447 RC-80740-465-2020, project 69553 as well as from the European Union’s Horizon 2020 research and innovation programme under the Marie Sklodowska-Curie grant agreement No 860881-HIDDeN. Some calculations were cross-checked with the Mathematica package xAct (www.xact.es).

\appendix
\section{Coefficients of the kinetic Lagrangian in \eq{lkin2}}\label{aA}
The coefficients of the kinetic  Lagrangian in \eq{lkin2}  involving only the vector field are given by
\begin{equation}\label{aterm}
 \mathcal{A}_{ab} \equiv 4(\gamma_{1}+\gamma_{2}) \delta_{ab} + 4(\omega_{1}+\omega_{3})\Bh_{a}^{\alpha}\Bh_{b\alpha}-4(\omega_{1}+\omega_{2}+\omega_{3}+
 \omega_{4}+\omega_{5})\Bs_{a}\Bs_{b} -4(\gamma_{3}+\gamma_{4})\Bs^{2} \delta_{ab} \,,
\end{equation}

\begin{align}\label{bterm}
 \mathcal{B}^{\sigma}_{ab} \equiv &(\kappa_{1}-\kappa_{3}+\omega_{2}+2\omega_{4})\Bs_{a}\Bh^{\sigma}_{b}+(\kappa_{2}+\kappa_{3}+\omega_{2}+2\omega_{5})\Bs_{b}\Bh^{\sigma}_{a} +(2\gamma_{3}+2\gamma_{4}+\chi_{1})\Bh^{\sigma}_{c}\Bs^{c}\delta_{ab} \notag \\
&-(\kappa_{4}+2\kappa_{5}-2\omega_{6})\Bh^{\alpha}_{a}\Bh^{\alpha_{1}}_{b}\varepsilon_{\alpha\alpha_{1}\alpha_{2}\alpha_{3}}h^{\sigma\alpha_{3}}n^{\alpha_{2}} \,,
\end{align}
\begin{align}\label{dterm}
\mathcal{D}^{\alpha\beta}_{ab} \equiv &
-(2\alpha_{1}+2\gamma_{2})h^{\alpha\beta}\delta_{ab}
 -(\beta_{5}+\beta_{6}+\kappa_{2}+\omega_{5})\Bh^{\alpha}_{b}\Bh^{\beta}_{a}\notag \\
 & -(\kappa_{1}+\omega_{4})\Bh^{\alpha}_{a}\Bh^{\beta}_{b}
 -(\alpha_{4}+\gamma_{4}+\chi_{1})\Bh^{\alpha}_{c}\Bh^{\beta c}\delta_{ab}  + 2(\beta_{7}-\omega_{3})\Bh^{\rho}_{a}\Bh_{b\rho}h^{\alpha\beta} \notag \\
 &+(\beta_{5}+\beta_{6}-2\beta_{7}-\kappa_{1}-\kappa_{2}+2\omega_{3}+\omega_{4}+\omega_{5})\Bs_{a}\Bs_{b}h^{\alpha\beta} +(\alpha_{4}+\gamma_{4}-\chi_{1})\Bs^{2} h^{\alpha\beta}\delta_{ab} \notag \\ &+(\beta_{1}+\kappa_{4}-\omega_{6})\Bh^{\rho}_{a}\Bh^{\alpha_{1}}_{b}\varepsilon_{\rho\alpha_{1}\alpha_{2}\alpha_{3}}h^{\alpha\alpha_{2}}h^{\beta\alpha_{3}} \notag \\
 &+(\beta_{1}-\beta_{2}+\beta_{3}+\kappa_{4}-\kappa_{6}+\kappa_{7}-\omega_{6})\Bs_{a}\Bh^{\rho}_{b}\varepsilon_{\rho\alpha_{1}\alpha_{2}\alpha_{3}}h^{\alpha\alpha_{2}}h^{\beta\alpha_{3}}n^{\alpha_{1}} \notag \\
 &-(\beta_{1}-\beta_{2}+\beta_{3}+\kappa_{4}-\kappa_{6}+\kappa_{7}-\omega_{6})\Bh^{\rho}_{a}\Bs_{b}\varepsilon_{\rho\alpha_{1}\alpha_{2}\alpha_{3}}h^{\alpha\alpha_{2}}h^{\beta\alpha_{3}}n^{\alpha_{1}} \,.
\end{align}
On the other hand, the coefficients of the terms mixing the vector field with the extrinsic curvature are given by
\begin{align}
\mathcal{C}^{\alpha\beta}_{a} \equiv & 2(\gamma_{3}-\omega_{2}-2\omega_{4})\Bh^{\alpha}_{c}\Bh^{\beta c}\Bs_{a}-2(\gamma_{3}+\gamma_{4}+\omega_{5})
 \Bh^{\alpha}_{c}\Bh^{\beta}_{a}\Bs^{c} -2(\gamma_{3}+\gamma_{4}+\omega_{5}) \Bh^{\alpha}_{a}\Bh^{\beta}_{c}\Bs^{c} \notag \\
 &+2(2\gamma_{1}+f_{X}) \Bs_{a} h^{\alpha\beta} -2( \gamma_{3}+2\omega_{1} +\omega_{2}) \Bs^{2} \Bs_{a} h^{\alpha\beta} +4\omega_{1}\Bh_{\rho}^{c}\Bh^{\rho}_{a} \Bs_{c} h^{\alpha\beta} \notag \\
 &-2\omega_{6}\Bh^{\rho}_{a}\Bh^{\alpha_{1}}_{c}\Bh^{\beta c}\varepsilon_{\rho\alpha_{1}\alpha_{2}\alpha_{3}}h^{\alpha\alpha_{3}}n^{\alpha_{2}} -2\omega_{6}\Bh^{\rho}_{a}\Bh^{\alpha_{1}}_{c}\Bh^{\alpha c}\varepsilon_{\rho\alpha_{1}\alpha_{2}\alpha_{3}}h^{\beta\alpha_{3}}n^{\alpha_{2}} \,,
\end{align}
\begin{align}
 \mathcal{E}^{\nu\alpha\beta}_{a} \equiv & \left(\gamma_{4}+\frac{1}{2}\kappa_{2}+\omega_{5}+\frac{1}{2}\chi_{1}\right)\Bh^{\nu}_{c}\Bh^{\alpha c}\Bh^{\beta}_a +\left(\gamma_{4}+\frac{1}{2}\kappa_{2}+\omega_{5}+\frac{1}{2}\chi_{1}\right)\Bh^{\nu}_{c}\Bh^{\alpha}_{a}\Bh^{\beta c} +(\kappa_{1}+2\omega_{4})\Bh^{\nu}_{a}\Bh^{\alpha}_{c}\Bh^{\beta c} \notag \\
 & +2\gamma_{2}\Bh^{\beta}_{a}h^{\nu\alpha}
 -\left(\gamma_{4} +\frac{1}{2}\kappa_{2} -\omega_{5} -\frac{1}{2}\chi_{1} \right)\Bs^{2}\Bh^{\beta}_{a}h^{\nu\alpha} \notag \\
 & +2\omega_{3}\Bh^{c}_{\lambda}\Bh^{\lambda}_{a}\Bh^{\beta}_{c}h^{\nu\alpha}+\left(\gamma_{4}+\frac{1}{2}\kappa_{2}-2\omega_{3}-\omega_{5}-\frac{1}{2}\chi_{1}\right)\Bh^{\beta}_{c}\Bs_{a}\Bs^{c}h^{\nu\alpha} \notag \\ 
 &+ 2\gamma_{2} \Bh^{\alpha}_{a}h^{\nu\beta}
 -\left(\gamma_{4} +\frac{1}{2}\kappa_{2}-\omega_{5}-\frac{1}{2}\chi_{1}\right)\Bs^{2} \Bh^{\alpha}_{a}h^{\nu\beta}  \notag \\
 & +2\omega_{3}\Bh^{c}_{\lambda}\Bh^{\lambda}_{a}\Bh^{\alpha}_{c}h^{\nu\beta}+\left(\gamma_{4}+\frac{1}{2}\kappa_{2}-2\omega_{3}-\omega_{5}-\frac{1}{2}\chi_{1}\right)\Bh^{\alpha}_{c}\Bs_{a}\Bs^{c} h^{\nu \beta} \notag \\
 & -(\kappa_{3}-\omega_{2})\Bs^{2}\Bh^{\nu}_{a}h^{\alpha\beta} +(2\gamma_{3}+\kappa_{3}+\omega_{2})\Bh^{\nu}_{c}\Bs_{a}\Bs^{c}h^{\alpha\beta} -2f_{X}\Bh^{\nu}_{a}h^{\alpha\beta} \notag \\ &-\left(\frac{1}{2}\kappa_{4}-\omega_{6}\right)\Bh^{\lambda}_{a}\Bh^{\alpha_{1}}_{c}\Bh^{\beta c}\varepsilon_{\lambda\alpha_{1}\alpha_{2}\alpha_{3}}h^{\nu\alpha_{2}}h^{\alpha\alpha_{3}} -\left(\frac{1}{2}\kappa_{4}-\omega_{6}\right)\Bh^{\lambda}_{a}\Bh^{\alpha_{1}}_{c}\Bh^{\alpha c}\varepsilon_{\lambda\alpha_{1}\alpha_{2}\alpha_{3}}h^{\nu\alpha_{2}}h^{\beta\alpha_{3}} \notag \\
 &-\left(\frac{1}{2}\kappa_{4}-\kappa_{6}-\omega_{6}+\chi_{2}\right)\Bh^{\lambda}_{c}\Bh^{\beta c}\Bs_{a}\varepsilon_{\lambda\alpha_{1}\alpha_{2}\alpha_{3}}h^{\nu\alpha_{2}}h^{\alpha\alpha_{3}}n^{\alpha_{1}} \notag \\
 & -(\kappa_{6}-\chi_{2})\Bh^{\lambda}_{c}\Bh^{\beta}_{a}\Bs^{c}\varepsilon_{\lambda\alpha_{1}\alpha_{2}\alpha_{3}}h^{\nu\alpha_{2}}h^{\alpha\alpha_{3}}n^{\alpha_{1}} \notag \\
 &+\left(\frac{1}{2}\kappa_{4}-\omega_{6}\right)\Bh^{\lambda}_{a}\Bh^{\beta}_{c}\Bs^{c}\varepsilon_{\lambda\alpha_{1}\alpha_{2}\alpha_{3}}h^{\nu\alpha_{2}}h^{\alpha\alpha_{3}}n^{\alpha_{1}} \notag \\
 & -\left(\frac{1}{2}\kappa_{4}-\kappa_{6}-\omega_{6}+\chi_{2}\right)\Bh^{\lambda}_{c}\Bh^{\alpha c}\Bs_{a}\varepsilon_{\lambda\alpha_{1}\alpha_{2}\alpha_{3}}h^{\nu\alpha_{2}}h^{\beta\alpha_{3}}n^{\alpha_{1}} \notag \\
 & -(\kappa_{6}-\chi_{2})\Bh^{\lambda}_{c}\Bh^{\alpha}_{a}\Bs^{c}\varepsilon_{\lambda\alpha_{1}\alpha_{2}\alpha_{3}}h^{\nu\alpha_{2}}h^{\beta\alpha_{3}}n^{\alpha_{1}} \notag \\
 & +\left(\frac{1}{2}\kappa_{4}-\omega_{6}\right)\Bh^{\lambda}_{a}\Bh^{\alpha}_{c}\Bs^{c}\varepsilon_{\lambda\alpha_{1}\alpha_{2}\alpha_{3}}h^{\nu\alpha_{2}}h^{\beta\alpha_{3}}n^{\alpha_{1}} \notag \\
 &+\left(\frac{1}{2}\kappa_{4}+\omega_{6}+2\kappa_{5}+\frac{1}{2}\kappa_{4}+\omega_{6}\right)\Bh^{\lambda}_{a}\Bh^{\alpha_{1}}_{c}\Bs^{c}\varepsilon_{\lambda\alpha_{1}\alpha_{2}\alpha_{3}}h^{\nu \alpha_{3}}h^{\alpha \beta}n^{\alpha_{2}} \,.
\end{align}
Finally, the coefficient of the term involving only the extrinsic curvature is given by
\begin{align}\label{fterm}
 \mathcal{F}^{\alpha\beta\rho\sigma} \equiv &-2(\gamma_{4}+\omega_{5})\Bh^{\alpha}_{c}\Bh^{\beta}_{b}\Bh^{\rho b}\Bh^{\sigma c}-2(\gamma_{4}+\omega_{5})\Bh^{\alpha}_{c}\Bh^{\beta}_{b}\Bh^{\rho c}\Bh^{\sigma b}-4\omega_{4}\Bh^{\alpha}_{c}\Bh^{\beta c}\Bh^{\rho}_{b}\Bh^{\sigma b} \notag \\
 & +2(\gamma_{3}-\omega_{2})\Bs{}^2 \Bh^{\rho}_{c}\Bh^{\sigma}_{c}h^{\alpha\beta}-4\gamma_{3}\Bh^{\rho}_{b}\Bs^{b}\Bh^{\sigma}_{c}\Bs^{c}h^{\alpha\beta}-2\gamma_{2}\Bh^{\beta}_{c}\Bh^{\sigma c} h^{\alpha\rho}+2(\gamma_{4}-\omega_{5} )\Bs^{2} \Bh^{\beta}_{c}\Bh^{\sigma c} h^{\alpha\rho} \notag \\
 & -2\omega_{3}\Bh^{\lambda}_{b}\Bh_{\lambda c}\Bh^{\beta b}\Bh^{\sigma c}h^{\alpha\rho}-2(\gamma_{4}-\omega_{3}-\omega_{5})\Bh^{\beta}_{c}\Bs^{c}\Bh^{\sigma}_{b}\Bs^{b}h^{\alpha\rho} \notag \\ 
 & -2\gamma_{2} \Bh^{\beta}_{c}\Bh^{\rho c} h^{\alpha\sigma}+2(\gamma_{4}-\omega_{5})\Bs^{2} \Bh^{\beta}_{c}\Bh^{\rho c} h^{\alpha\sigma}-2\omega_{3}\Bh^{b}_\lambda \Bh^{\lambda}_{c}\Bh^{\beta c}\Bh^{\rho}_{b}h^{\alpha\sigma} \notag \\
 & -(\gamma_{4}-\omega_{3}+\omega_{5})\Bh^{\beta}_{c}\Bs^{c}\Bh^{\rho}_{b}\Bs^{b}h^{\alpha\sigma}-2\gamma_{2}\Bh^{\alpha}_{c}\Bh^{\sigma c} h^{\beta\rho}+2(\gamma_{4}-\omega_{5})\Bs^{2}\Bh^{\alpha}_{c}\Bh^{\sigma c} h^{\beta\rho} \notag \\
 &-2\omega_{3}\Bh_{\lambda b}\Bh^{\lambda}_{c}\Bh^{\alpha c} \Bh^{\sigma b} h^{\beta\rho}-2(\gamma_{4}-\omega_{3}-\omega_{5})\Bh^{\alpha}_{c}\Bh^{\sigma}_{b}\Bs^{c}\Bs^{b}h^{\beta\rho} \notag \\
 &+2\gamma_{2}\Bs^{2}h^{\alpha\sigma}h^{\beta\rho}-2\omega_{3}\Bs^{4}h^{\alpha\sigma}h^{\beta\rho}+2\omega_{3}\Bh_{\lambda b}\Bh^{\lambda}_{c}\Bs^{b}\Bs^{c}h^{\alpha\sigma}h^{\beta\rho}+\frac{1}{2}f h^{\alpha\sigma}h^{\beta\rho} \notag \\
 & -2\gamma_{2}\Bh^{\alpha}_{c}\Bh^{\rho c} h^{\beta\sigma}+2(\gamma_{4}-\omega_{5})\Bs^{2}\Bh^{\alpha}_{c}\Bh^{\rho c} h^{\beta\sigma}-2\omega_{3}\Bh_{\lambda b}\Bh^{\lambda}_{c}\Bh^{\alpha c}\Bh^{\rho b} h^{\beta\sigma} \notag \\
 & -2(\gamma_{4}-\omega_{3}-\omega_{5})\Bh^{\alpha}_{c}\Bh^{\rho}_{b}\Bs^{c}\Bs^{b} h^{\beta\sigma} \notag \\ &+2\gamma_{2}\Bs^{2} h^{\alpha\rho}h^{\beta\sigma}-2 \omega_{3}\Bs^{4} h^{\alpha\rho}h^{\beta\sigma}+2\omega_{3}\Bh_{\lambda b}\Bh^{\lambda}_{c}\Bs^{b}\Bs^{c} h^{\alpha\rho}h^{\beta\sigma}+\frac{1}{2}f h^{\alpha\rho}h^{\beta\sigma}  \notag \\
 & +2(\gamma_{3} - \omega_2)\Bs^{2}\Bh^{\alpha}_{c}\Bh^{\beta c} h^{\rho\sigma}-4\gamma_{3}\Bh^{\alpha}_{c}\Bh^{\beta}_{b}\Bs^{c}\Bs^{b} h^{\rho\sigma} +2 f_{X} \Bh^{\rho}_{c}\Bh^{\sigma c} h^{\alpha\beta}+2 f_{X} \Bh^{\alpha}_{c}\Bh^{\beta c} h^{\rho\sigma}  \notag \\
 & + 4 \gamma_{1}\Bs^{2} h^{\alpha\beta}h^{\rho\sigma}-4\omega_{1}\Bs^{4}h^{\alpha\beta}h^{\rho\sigma}+4\omega_{1}\Bh_{b\lambda}\Bh^{\lambda}_{c}\Bs^{b}\Bs^{c}h^{\alpha\beta}h^{\rho\sigma}-f h^{\alpha\beta}h^{\rho\sigma} \notag \\
 & -\omega_{6}\Bh^{\lambda}_{c}\Bh^{\alpha_{1}}_{b}\Bh^{\beta c}\Bh^{\sigma b}\varepsilon_{\lambda\alpha_{1}\alpha_{2}\alpha_{3}}h^{\alpha\alpha_{2}}h^{\rho\alpha_{3}} -\omega_{6}\Bh^{\lambda}_{c}\Bh^{\alpha_{1}}_{b}\Bh^{\alpha c}\Bh^{\sigma b}\varepsilon_{\lambda\alpha_{1}\alpha_{2}\alpha_{3}}h^{\beta\alpha_{2}}h^{\rho\alpha_{3}} \notag \\
 &-\omega_{6}\Bh^{\lambda}_{c}\Bh^{\alpha_{1}}_{b}\Bh^{\beta c}\Bh^{\rho b}\varepsilon_{\lambda\alpha_{1}\alpha_{2}\alpha_{3}}h^{\alpha\alpha_{2}}h^{\sigma\alpha_{3}}
 -\omega_{6}\Bh^{\lambda}_{c}\Bh^{\alpha_{1}}_{b}\Bh^{\alpha c}\Bh^{\rho b}\varepsilon_{\lambda\alpha_{1}\alpha_{2}\alpha_{3}}h^{\beta\alpha_{2}}h^{\sigma\alpha_{3}} \notag \\
 & -\omega_{6}\Bh^{\lambda}_{b}\Bh^{\beta}_{c}\Bh^{\sigma b}\Bs^{c}\varepsilon_{\lambda\alpha_{1}\alpha_{2}\alpha_{3}}h^{\alpha\alpha_{2}}h^{\rho\alpha_{3}}n^{\alpha_{1}}+\omega_{6}\Bh^{\lambda}_{c}\Bh^{\beta c}\Bh^{\sigma}_{b}\Bs^{b}\varepsilon_{\lambda\alpha_{1}\alpha_{2}\alpha_{3}}h^{\alpha\alpha_{2}}h^{\rho\alpha_{3}}n^{\alpha_{1}} \notag \\
 & -\omega_{6}\Bh^{\lambda}_{b}\Bh^{\alpha}_{c}\Bh^{\sigma b}\Bs^{c}\varepsilon_{\lambda\alpha_{1}\alpha_{2}\alpha_{3}}h^{\beta\alpha_{2}}h^{\rho\alpha_{3}}n^{\alpha_{1}}+\omega_{6}\Bh^{\lambda}_{c}\Bh^{\alpha c}\Bh^{\sigma}_{b}\Bs^{b}\varepsilon_{\lambda\alpha_{1}\alpha_{2}\alpha_{3}}h^{\beta\alpha_{2}}h^{\rho\alpha_{3}}n^{\alpha_{1}} \notag \\
 & -\omega_{6}\Bh^{\lambda}_{b}\Bh^{\beta}_{c}\Bh^{\rho b} \Bs^{c}\varepsilon_{\lambda\alpha_{1}\alpha_{2}\alpha_{3}}h^{\alpha\alpha_{2}}h^{\sigma\alpha_{3}}n^{\alpha_{1}}+\omega_{6}\Bh^{\lambda}_{c}\Bh^{\beta c} \Bh^{\rho}_{b}\Bs^{b}\varepsilon_{\lambda\alpha_{1}\alpha_{2}\alpha_{3}}h^{\alpha\alpha_{2}}h^{\sigma\alpha_{3}}n^{\alpha_{1}} \notag \\
 & -\omega_{6}\Bh^{\lambda}_{b}\Bh^{\alpha}_{c}\Bh^{\rho b} \Bs^{c}\varepsilon_{\lambda\alpha_{1}\alpha_{2}\alpha_{3}}h^{\beta\alpha_{2}}h^{\sigma\alpha_{3}}n^{\alpha_{1}}+\omega_{6}\Bh^{\lambda}_{c}\Bh^{\alpha c} \Bh^{\rho}_{b}\Bs^{b}\varepsilon_{\lambda\alpha_{1}\alpha_{2}\alpha_{3}}h^{\beta\alpha_{2}}h^{\sigma\alpha_{3}}n^{\alpha_{1}} \notag \\
 & -\omega_{6}\Bh^{\lambda}_{c}\Bh^{\alpha_{1}}_{b}\Bh^{\sigma c} \Bs^{b}\varepsilon_{\lambda\alpha_{1}\alpha_{2}\alpha_{3}}h^{\alpha\rho}h^{\beta\alpha_{3}}n^{\alpha_{2}}-\omega_{6}\Bh^{\lambda}_{c}\Bh^{\alpha_{1}}_{b}\Bh^{\rho c} \Bs^{b}\varepsilon_{\lambda\alpha_{1}\alpha_{2}\alpha_{3}}h^{\alpha\sigma}h^{\beta\alpha_{3}}n^{\alpha_{2}} \notag \\
 & -\omega_{6}\Bh^{\lambda}_{c}\Bh^{\alpha_{1}}_{b}\Bh^{\sigma c} \Bs^{b}\varepsilon_{\lambda\alpha_{1}\alpha_{2}\alpha_{3}}h^{\alpha\alpha_{3}}h^{\beta\rho}n^{\alpha_{2}}-\omega_{6}\Bh^{\lambda}_{c}\Bh^{\alpha_{1}}_{b}\Bh^{\rho c} \Bs^{b}\varepsilon_{\lambda\alpha_{1}\alpha_{2}\alpha_{3}}h^{\alpha\alpha_{3}}h^{\beta\sigma}n^{\alpha_{2}} \notag \\ 
 & -\omega_{6}\Bh^{\lambda}_{c}\Bh^{\alpha_{1}}_{b}\Bh^{\beta c} \Bs^{b}\varepsilon_{\lambda\alpha_{1}\alpha_{2}\alpha_{3}}h^{\alpha\sigma}h^{\rho\alpha_{3}}n^{\alpha_{2}} - \omega_{6}\Bh^{\lambda}_{c}\Bh^{\alpha_{1}}_{b}\Bh^{\alpha c} \Bs^{b}\varepsilon_{\lambda\alpha_{1}\alpha_{2}\alpha_{3}}h^{\beta\sigma}h^{\rho\alpha_{3}}n^{\alpha_{2}} \notag \\
 &-\omega_{6}\Bh^{\lambda}_{c}\Bh^{\alpha_{1}}_{b}\Bh^{\beta c} \Bs^{b}\varepsilon_{\lambda\alpha_{1}\alpha_{2}\alpha_{3}}h^{\alpha\rho}h^{\sigma\alpha_{3}}n^{\alpha_{2}}-\omega_{6}\Bh^{\lambda}_{c}\Bh^{\alpha_{1}}_{b}\Bh^{\alpha c} \Bs^{b}\varepsilon_{\lambda\alpha_{1}\alpha_{2}\alpha_{3}}h^{\beta\rho}h^{\sigma\alpha_{3}}n^{\alpha_{2}} \,.
\end{align}
\section{Coefficients of the kinetic Lagrangian in \eq{lkinbasis}}\label{aB}
The coefficients of the kinetic Lagrangian in \eq{lkinbasis} 
are given by
\begin{subequations}\label{atermbasis}
\begin{align}
\A_1 &\equiv 4\left(\gamma_{1}+\gamma_{2}-(\gamma_{3} +\gamma_{4} + \omega_1 + \omega_2 + \omega_3 + \omega_4 + \omega_5)\Bs^{2}+(\omega_{1}+\omega_3) (\hat{\tilde{B}}_{1}^{\ 1})^{2} \right) \,,\\
\A_2 &\equiv 4\left(\gamma_{1}+\gamma_{2}-(\gamma_{3}+\gamma_{4})\Bs^{2}+(\omega_{1}+\omega_{3})(\hat{\tilde{B}}_{1}^{\ 2})^{2}\right) \,,\\
\A_3 &\equiv 4\left(\gamma_{1}+\gamma_{2}-(\gamma_{3}+\gamma_{4})\Bs^{2}+(\omega_{1}+\omega_3) \Bh^{2}-(\omega_{1}+\omega_3)(\hat{\tilde{B}}_{1}^{\ 1})^{2}-(\omega_{1}+\omega_3) (\hat{\tilde{B}}_{1}^{\ 2})^{2} \right) \,,\\
\A_4 &\equiv 4 (\omega_{1}+\omega_{3}) \ \hat{\tilde{B}}_{1}^{\ 1}\hat{\tilde{B}}_{1}^{\ 2} \,,\\
\A_5 &\equiv 4 (\omega_{1}+\omega_{3}) \ \hat{\tilde{B}}_{1}^{\ 1}\sqrt{\Bh^{2}-(\hat{\tilde{B}}_{1}^{\ 1})^{2}-(\hat{\tilde{B}}_{1}^{\ 2})^{2}} \,,\\
\A_6 &\equiv 4 (\omega_{1}+\omega_{3}) \ \hat{\tilde{B}}_{1}^{\ 2}\sqrt{\Bh^{2}-(\hat{\tilde{B}}_{1}^{\ 1})^{2}-(\hat{\tilde{B}}_{1}^{\ 2})^{2}} \,,\\
\B_1 &\equiv  (2\gamma_{3}+2\gamma_{4}+\kappa_{1}+\kappa_{2}+2\omega_{2}+2\omega_{4}+2\omega_{5}+\chi_{1}) \ \hat{\tilde{B}}_{1}^{\ 1}\sqrt{\Bs^{2}} \,,\\
\B_2 &\equiv  (\kappa_{1}-\kappa_{3}+\omega_{2}+2\omega_{4}) \ \hat{\tilde{B}}_{1}^{\ 2}\sqrt{\Bs^{2}} \,,\\
\B_3 &\equiv  (\kappa_{1}-\kappa_{3}+\omega_{2}+2\omega_{4}) \ \sqrt{\Bh^{2}-(\hat{\tilde{B}}_{1}^{\ 1})^{2}-(\hat{\tilde{B}}_{1}^{\ 2})^{2}}\ \sqrt{\Bs^{2}}  \,,\\
\B_4 &\equiv  (\kappa_{2}+\kappa_{3}+\omega_{2}+2\omega_{5}) \ \hat{\tilde{B}}_{1}^{\ 2}\sqrt{\Bs^{2}} \,,\\
\B_5 &\equiv  (\kappa_{2}+\kappa_{3}+\omega_{2}+2\omega_{5}) \ \sqrt{\Bh^{2}-(\hat{\tilde{B}}_{1}^{\ 1})^{2}-(\hat{\tilde{B}}_{1}^{\ 2})^{2}}\ \sqrt{\Bs^{2}} \,,\\
\B_6 &\equiv  (2\gamma_{3}+2\gamma_{4}+\chi_{1}) \ \hat{\tilde{B}}_{1}^{\ 1}\sqrt{\Bs^{2}} \,,\\
\C_1 &\equiv 2\left(2\gamma_{1}+(\gamma_{3}-\omega_2-2\omega_4) \Bh^{2}-2(\gamma_{3}+\gamma_4-\omega_1+\omega_5)(\hat{\tilde{B}}_{1}^{\ 1})^{2}-(\gamma_{3}+2\omega_1+\omega_2) \Bs^{2}+f_{X} \right)  \sqrt{\Bs^{2}} \,,\\
\C_2&\equiv 2\left(2\gamma_{1}-(\gamma_{3}+2\omega_1+\omega_2) \Bs^{2}+2\omega_{1}(\hat{\tilde{B}}_{1}^{\ 1})^{2}+f_{X}\right)  \sqrt{2\Bs^{2}} \,, \\
\C_3&\equiv -4(\gamma_{3}+\gamma_{4}-\omega_{1}+\omega_{5}) \ \hat{\tilde{B}}_{1}^{\ 1}\hat{\tilde{B}}_{1}^{\ 2}\sqrt{\Bs^{2}} \,, \\
\C_4&\equiv 4 \omega_{1}  \hat{\tilde{B}}_{1}^{\ 1}\hat{\tilde{B}}_{1}^{\ 2} \sqrt{2\Bs^{2}} \,, \\
\C_5&\equiv -4 (\gamma_{3}+\gamma_{4}-\omega_{1}+\omega_{5}) \ \hat{\tilde{B}}_{1}^{\ 1}\sqrt{\Bh^{2}-(\hat{\tilde{B}}_{1}^{\ 1})^{2}-(\hat{\tilde{B}}_{1}^{\ 2})^{2}}\ \sqrt{\Bs^{2}} \,, \\
\C_6&\equiv 4 \omega_{1}  \sqrt{\Bh^{2}-(\hat{\tilde{B}}_{1}^{\ 1})^{2}-(\hat{\tilde{B}}_{1}^{\ 2})^{2}} \sqrt{2\Bs^{2}} \,,
\end{align}
\end{subequations}
\begin{subequations}
\begin{align}
\D_1&\equiv -2(\alpha_{1}+\gamma_2)+(\alpha_{4}+\beta_5+\beta_6-2\beta_7+\gamma_4-\kappa_1-\kappa_2+2\omega_3+\omega_4+\omega_5-\chi_1) \Bs^{2} -(\alpha_{4}+\gamma_4+\chi_1)\Bh^{2} \notag \\
&-(\beta_{5}+\beta_6-2\beta_7+\kappa_1+\kappa_2+2\omega_3+\omega_4+\omega_5) (\hat{\tilde{B}}_{1}^{\ 1})^{2} \,, \\
\D_2&\equiv -2(\alpha_{1}-\gamma_1)+(\alpha_{4}+\gamma_4-\chi_1)\Bs^{2}-(\alpha_{4}+\gamma_4-\chi_1)\Bh^{2} \notag \\
&-(\beta_{5}+\beta_6-2\beta_7+\kappa_1+\kappa_2+2\omega_3+\omega_4+\omega_5)(\hat{\tilde{B}}_{1}^{\ 2})^{2} \,, \\
\D_3&\equiv -2(\alpha_{1}+\gamma_2)+(\alpha_{4} + \gamma_4 - \chi_1) \Bs^{2} -(\alpha_{4} + \beta_5 + \beta_6 - 2\beta_7 + \gamma_4 +\kappa_1 +\kappa_2 +2\omega_3 + \omega_4 + \omega_5 + \chi_1) \Bh^{2} \notag \\
& +(\beta_{5} + \beta_6 - 2\beta_7 + \kappa_1 + \kappa_2 + 2\omega_3 + \omega_4 + \omega_5)(\hat{\tilde{B}}_{1}^{\ 1})^{2} \notag \\
& +(\beta_{5} + \beta_6 - 2\beta_7 + \kappa_1 + \kappa_2 + 2\omega_3 + \omega_4 + \omega_5)(\hat{\tilde{B}}_{1}^{\ 2})^{2} \,,\\
\D_4& \equiv \left(\beta_{5}+\beta_{6}-2\beta_{7}+\kappa_{1}+\kappa_{2}+2\omega_{3}+\omega_{4}+\omega_{5} \right)\hat{\tilde{B}}_{1}^{\ 1} \hat{\tilde{B}}_{1}^{\ 2} \,, \\
\D_5& \equiv \left(\beta_{5}+\beta_{6}-2\beta_{7}+\kappa_{1}+\kappa_{2}+2\omega_{3}+\omega_{4}+\omega_{5} \right) \hat{\tilde{B}}_{1}^{\ 1} \sqrt{\Bh^{2}-(\hat{\tilde{B}}_{1}^{\ 1})^{2}-(\hat{\tilde{B}}_{1}^{\ 2})^{2}} \,,\\
\D_6& \equiv -\left(\beta_{5}+\beta_{6}-2\beta_{7}+\kappa_{1}+\kappa_{2}+2\omega_{3}+\omega_{4}+\omega_{5} \right) \hat{\tilde{B}}_{1}^{\ 2} \sqrt{\Bh^{2}-(\hat{\tilde{B}}_{1}^{\ 1})^{2}-(\hat{\tilde{B}}_{1}^{\ 2})^{2}} \,,\\
\D_7& \equiv -2(\alpha_{1}+\gamma_{2})+\left(\alpha_{4}+\beta_{5}+\beta_{6}-2\beta_{7}+\gamma_{4}-\kappa_{1}-\kappa_{2}+2\omega_{3}+\omega_{4}+\omega_{5}-\chi_{1}\right)\Bs^{2}+2(\beta_{7}-\omega_{3})(\hat{\tilde{B}}_{1}^{\ 1})^{2} \,,\\
\D_8&\equiv -2(\alpha_{1}+\gamma_{2})+(\alpha_{4}+\gamma_{4}-\chi_{1})\Bs^{2} +2(\beta_{7}-\omega_{3})(\hat{\tilde{B}}_{1}^{\ 2})^{2} \,,\\
\D_9&\equiv -2(\alpha_{1}+\gamma_{2})+(\alpha_{4}+\gamma_{4}-\chi_{1})\Bs^{2}+2(\beta_{7}-\omega_{3})\left[\Bh^{2}-(\hat{\tilde{B}}_{1}^{\ 1})^{2}-(\hat{\tilde{B}}_{1}^{\ 2})^{2}\right] \,,\\
\D_{10}&\equiv -2(\alpha_{1}+\gamma_{2})+(\alpha_{4}+\beta_{5}+\beta_{6}-2\beta_{7}+\gamma_{4}-\kappa_{1}-\kappa_{2}+2\omega_{3}+\omega_{4}+\omega_{5}-\chi_{1})\Bs^{2}+2(\beta_{7}-\omega_{3})(\hat{\tilde{B}}_{1}^{\ 1})^{2} \,,\\
\D_{11}&\equiv -2(\alpha_{1}+\gamma_{2})+(\alpha_{4}+\gamma_{4}-\chi_{1})\Bs^{2}+2(\beta_{7}-\omega_{3})(\hat{\tilde{B}}_{1}^{\ 2})^{2},\\
\D_{12}&\equiv -2(\alpha_{1}+\gamma_{2})
+(\alpha_{4}+\gamma_{4}-\chi_{1})\Bs^{2}+2(\beta_{7}-\omega_{3})\left[\Bh^{2}-(\hat{\tilde{B}}_{1}^{\ 1})^{2}-(\hat{\tilde{B}}_{1}^{\ 2})^{2}\right] \,,\\
\D_{13}&\equiv 2(\beta_{7}-\omega_{3})\hat{\tilde{B}}_{1}^{\ 1} \hat{\tilde{B}}_{1}^{\ 2} \,,\\
\D_{14}&\equiv 2(\beta_{7}-\omega_{3})\hat{\tilde{B}}_{1}^{\ 1} \sqrt{\Bh^{2}-(\hat{\tilde{B}}_{1}^{\ 1})^{2}-(\hat{\tilde{B}}_{1}^{\ 2})^{2}} \,,\\
\D_{15}&\equiv -(\beta_{1}-\beta_{2}+\beta_{3}+\kappa_{4}-\kappa_{6}+\kappa_{7}-\omega_{6})\hat{\tilde{B}}_{1}^{\ 2} \sqrt{\Bs^{2}} \,,\\
\D_{16}&\equiv -(\beta_{1}-\beta_{2}+\beta_{3}+\kappa_{4}-\kappa_{6}+\kappa_{7}-\omega_{6})\sqrt{\Bh^{2}-(\hat{\tilde{B}}_{1}^{\ 1})^{2}-(\hat{\tilde{B}}_{1}^{\ 2})^{2}}\sqrt{\Bs^{2}} \,,\\
\D_{17}&\equiv 2(\beta_{7}-\omega_{3})\hat{\tilde{B}}_{1}^{\ 2}\sqrt{\Bh^{2}-(\hat{\tilde{B}}_{1}^{\ 1})^{2}-(\hat{\tilde{B}}_{1}^{\ 2})^{2}} \,,\\
\D_{18}&\equiv (\beta_{1}-\beta_{2}+\beta_{3}+\kappa_{4}-\kappa_{6}+\kappa_{7}-\omega_{6})\hat{\tilde{B}}_{1}^{\ 2} \sqrt{\Bs^{2}} \,,\\
\D_{19}&\equiv (\beta_{1}-\beta_{2}+\beta_{3}+\kappa_{4}-\kappa_{6}+\kappa_{7}-\omega_{6})\sqrt{\Bh^{2}-(\hat{\tilde{B}}_{1}^{\ 1})^{2}-(\hat{\tilde{B}}_{1}^{\ 2})^{2}}\sqrt{\Bs^{2}} \,,\\
\D_{20}&\equiv 2(\beta_{7}-\omega_{3})\hat{\tilde{B}}_{1}^{\ 1} \hat{\tilde{B}}_{1}^{\ 2} \,,\\
\D_{21}&\equiv 2(\beta_{7}-\omega_{3})\hat{\tilde{B}}_{1}^{\ 1} \sqrt{\Bh^{2}-(\hat{\tilde{B}}_{1}^{\ 1})^{2}-(\hat{\tilde{B}}_{1}^{\ 2})^{2}} \,,\\
\D_{22}&\equiv 2(\beta_{7}-\omega_{3})\hat{\tilde{B}}_{1}^{\ 2}\sqrt{\Bh^{2}-(\hat{\tilde{B}}_{1}^{\ 1})^{2}-(\hat{\tilde{B}}_{1}^{\ 2})^{2}} \,,
\end{align}
\end{subequations}
\begin{subequations}\label{ftermbasis}
\begin{align}
\E_1&\equiv \left(4\gamma_{2}+2(\gamma_{3}+\omega_{2}-4\omega_{3}) \Bs^{2} +(2\gamma_{4}+\kappa_{1}+\kappa_{2}+4\omega_{3}+2\omega_{4}+2\omega_{5}+\chi_{1})\Bh^{2} -2f_{X}\right) \hat{\tilde{B}}_{1}^{\ 1} \,, \\
\E_2&\equiv 2\sqrt{2}\left((\gamma_{3}+\omega_{2})\Bs^{2}-f_{X}\right) \hat{\tilde{B}}_{1}^{\ 1} \,,\\
\E_3&\equiv \left(4\gamma_{2}+(-2\gamma_{4}-\kappa_{2}-\kappa_{3}+\omega_{2} +2\omega_{5}+\chi_{1} )\Bs^{2} + (2\gamma_{4}+\kappa_{1}+\kappa_{2}+4\omega_{3}+2\omega_{4}+2\omega_{5}+\chi_{1})\Bh^{2}-2f_{X} \right) \hat{\tilde{B}}_{1}^{\ 2} \,, \\
\E_4&\equiv -\sqrt{2}\left((\kappa_{3}-\omega_{2})\Bs^{2}+2f_{X}\right)\hat{\tilde{B}}_{1}^{\ 2} \,,\\
\E_5&\equiv \Bigl(4\gamma_{2}-(2\gamma_{4}+\kappa_{2}+\kappa_{3}-\omega_{2}-2\omega_{5}-\chi_{1})\Bs^{2} \notag +\\
& (2\gamma_{4}+\kappa_{1}+\kappa_{2}+4\omega_{3}+2\omega_{4}+2\omega_{5}+\chi_{1})\Bh^{2} -2f_{X}
\Bigr) \sqrt{\Bh^{2}-(\hat{\tilde{B}}_{1}^{\ 1})^{2}-(\hat{\tilde{B}}_{1}^{\ 2})^{2}} \,, \\
\E_6&\equiv \left((\omega_{2}-\kappa_{3})\Bs^{2}-2f_{X}\right) \sqrt{2\left(\Bh^{2}-(\hat{\tilde{B}}_{1}^{\ 2})^{2}-(\hat{\tilde{B}}_{1}^{\ 1})^{2}\right)} \,,\\
\E_7&\equiv 2\sqrt{2}\left(\gamma_{2}+\omega_3 \Bh^{2}-\omega_3 \Bs^{2}\right) \hat{\tilde{B}}_{1}^{\ 1} \,,\\
\E_8&\equiv \frac{1}{\sqrt{2}}(\kappa_{4}-2\kappa_{6}-2\omega_{6}+2\chi_{2}) \sqrt{\Bs^{2}}\left(\Bh^{2}-(\hat{\tilde{B}}_{1}^{\ 1})^{2}\right) \,,\\
\E_{9}&\equiv \frac{1}{\sqrt{2}}\left(4\gamma_{2}+4\omega_{3}\Bh^{2}-(2\gamma_{4}+\kappa_{2}-2\omega_{5}-\chi_{1})\Bs^{2}\right) (\hat{\tilde{B}}_{1}^{\ 2}) \,,\\
\E_{10}&\equiv -\frac{1}{\sqrt{2}}(\kappa_{4}-2\kappa_{6}-2\omega_{6}+2\chi_{2}) \hat{\tilde{B}}_{1}^{\ 1} \hat{\tilde{B}}_{1}^{\ 2} \sqrt{\Bs^{2}} \,,\\
\E_{11}&\equiv \frac{1}{\sqrt{2}}\left(4\gamma_{2}+4\omega_{3}\Bh^{2}-(2\gamma_{4}+\kappa_{2}-2\omega_{5}-\chi_{1})\Bs^{2}\right) \sqrt{\Bh^{2}-(\hat{\tilde{B}}_{1}^{\ 1})^{2}-(\hat{\tilde{B}}_{1}^{\ 2})^{2}} \,,\\
\E_{12}&\equiv -\frac{1}{\sqrt{2}}(\kappa_{4}-2\kappa_{6}-2\omega_{6}+2\chi_{2})\hat{\tilde{B}}_{1}^{\ 1} \sqrt{\Bs^{2}}\sqrt{\Bh^{2}-(\hat{\tilde{B}}_{1}^{\ 1})^{2}-(\hat{\tilde{B}}_{1}^{\ 2})^{2}} \,,\\
\E_{13}&\equiv -\frac{1}{\sqrt{2}}(\kappa_{4}-2\kappa_{6}-2\omega_{6}+2\chi_{2}) \left(\Bh^{2}-(\hat{\tilde{B}}_{1}^{\ 1})^{2}\right)\sqrt{\Bs^{2}} \,,\\
\E_{14}&\equiv 2\sqrt{2}\left(\gamma_{2}+\omega_3\Bh^{2}-\omega_3\Bs^{2}\right) \hat{\tilde{B}}_{1}^{\ 1} \,,\\
\E_{15}&\equiv \frac{1}{\sqrt{2}}(\kappa_{4}-2\kappa_{6}-2\omega_{6}+2\chi_{2}) \hat{\tilde{B}}_{1}^{\ 1} \hat{\tilde{B}}_{1}^{\ 2}\sqrt{\Bs^{2}} \,,\\
\E_{16}&\equiv \frac{1}{\sqrt{2}}\left(4\gamma_{2}+4\omega_{3}\Bh^{2}-(2\gamma_{4}+\kappa_{2}-2\omega_{5}-\chi_{1})\Bs^{2} \right) \hat{\tilde{B}}_{1}^{\ 2} \,,\\
\E_{17}&\equiv \frac{1}{\sqrt{2}}(\kappa_{4}-2\kappa_{6}-2\omega_{6}+2\chi_{2}) \hat{\tilde{B}}_{1}^{\ 1}\sqrt{\Bh^{2}-(\hat{\tilde{B}}_{1}^{\ 1})^{2}-(\hat{\tilde{B}}_{1}^{\ 2})^{2}}\sqrt{\Bs^{2}} \,,\\
\E_{18}&\equiv \frac{1}{\sqrt{2}}\left(4\gamma_{2}+4\omega_{3}\Bh^{2}-(2\gamma_{4}+\kappa_{2}-2\omega_{5}-\chi_{1}) \Bs^{2} \right) \sqrt{\Bh^{2}-(\hat{\tilde{B}}_{1}^{\ 1})^{2}-(\hat{\tilde{B}}_{1}^{\ 2})^{2}} \,,\\
\F_{1}&\equiv 4\left(\gamma_{1}+\gamma_{2}-(2\gamma_2-\gamma_{3}-2\gamma_4+\omega_2+2\omega_5)\Bh^{2}-(2\gamma_{3}+2\gamma_4-\omega_1-3\omega_3-2\omega_5)(\hat{\tilde{B}}_{1}^{\ 1})^{2} \right. \notag \\
&\left. -(\omega_{1}+\omega_{3})\Bs^{4}-(\gamma_{4}+\omega_{3}+\omega_{4}+\omega_{5})\Bh^{4}+f_{X}\Bh^{2} \right) \,,\\
\F_{2}&\equiv 4\left(2\gamma_{1}+\gamma_{2}+(2\omega_{1}+\omega_{3})\left(2(\hat{\tilde{B}}_{1}^{\ 1})^{2}-\Bs^{2}\right) \right)\Bs^{2}-f(X) \,,\\
\F_{3}&\equiv \sqrt{2}\Biggl[2 \Bigg(2\gamma_{1}-2\left((\gamma_{3}-\omega_{1})(\hat{\tilde{B}}_{1}^{\ 1})^{2}+\omega_{1}\Bs^{2}\right)+(\gamma_{3}-\omega_2)\Bh^{2}\Biggr)\Bs^{2} -f+2f_{X}\Bh^{2}\Biggr] \,, \\
\F_{4}&\equiv 4\left((\gamma_2-\omega_{3})\Bh^{2}+\left(\gamma_{2}-\omega_{3}\Bs^{2}-(\gamma_{4}-2\omega_{3}-\omega_{5})(\hat{\tilde{B}}^{1}_{\ 1})^{2}+(\gamma_4-\omega_5)\Bh^2\right)\Bs^{2}\right)+f \,, \\
\F_{5}&\equiv \left(\gamma_{2}+\omega_{3}(\hat{\tilde{B}}_{1}^{\ 1})^{2}-\omega_{3}\Bs^{2}\right) \Bs^2 +f \,.
\end{align}
\end{subequations}
\section{Coefficients of the kinetic Lagrangian in \eq{eq:kinm}}\label{aC}
The kinetic matrix entries in \eq{eq:kinm} are given by
\begin{equation}\label{atermscalar}
    \tilde{\mathcal{A}}_{ab}\equiv 4\left((\gamma_{1}+\gamma_{2})\delta_{ab}+(\omega_{1}+\omega_{3})\Zh_{a}^{\alpha}\Zh_{\alpha b}-(\omega_{1}+\omega_{2}+\omega_{3}+\omega_{4}+\omega_{5})\Zs_{a}\Zs_{b}-(\gamma_{3}+\gamma_{4})\Zs^{2}\delta_{ab} \right) \,,
\end{equation}
\begin{align}
    \tilde{\mathcal{C}}^{\rho\sigma}_{a}\equiv &4(\gamma_{3}-\omega_{2}-2\omega_{4})\Zh^{\rho}_{c}\Zh^{\sigma c}\Zs_{a}-4(\gamma_{3}+\gamma_{4}+\omega_{5})\Zh^{\rho}_{c}\Zh^{\sigma}_{a}\Zs^{c}-4(\gamma_{3}+\gamma_{4}+\omega_{5})\Zh^{\rho}_{a}\Zh^{\sigma}_{c}\Zs^{c} \notag \\
    & +8\gamma_{1} \Zs_{a} h^{\rho\sigma} -4 (\gamma_{3}+2\omega_{1}+\omega_{2})\Zs^{2} \Zs_{a} h^{\rho\sigma} +4 f_{X} \Zs_{a} h^{\rho\sigma} +8\omega_{1}\Zh_{\lambda c}\Zh^{\lambda}_{a} 
    \Zs^c h^{\rho\sigma} \\
    & -4\omega_{6}\Zh^{\lambda}_{a}\Zh^{\alpha_{1}}_{c}\Zh^{\sigma c}\varepsilon_{\lambda\alpha_{1}\alpha_{2}\alpha_{3}}h^{\rho\alpha_{3}}n^{\alpha_{2}}-4\omega_{6}\Zh^{\lambda}_{a}\Zh^{\alpha_{1}}_{c}\Zh^{\rho c}\varepsilon_{\lambda\alpha_{1}\alpha_{2}\alpha_{3}}h^{\sigma\alpha_{3}}n^{\alpha_{2}} \,, \notag
\end{align}
\begin{align}\label{ftermscalar}
    \tilde{\mathcal{F}}^{\alpha\beta\rho\sigma}\equiv &-2(\gamma_{4}+\omega_{5})\Zh^{\alpha}_{a}\Zh^{\beta}_{b}\Zh^{\rho b}\Zh^{\sigma a}-2(\gamma_{4}+\omega_{5})\Zh^{\alpha}_{a}\Zh^{\beta}_{b}\Zh^{\rho a}\Zh^{\sigma b}-4\omega_{4}\Zh^{\alpha}_{a}\Zh^{\beta a} \Zh^{\rho}_{b}\Zh^{\sigma b}  +2(\gamma_{3}-\omega_{2})\Zs^{2}\Zh^{\rho}_{c}\Zh^{\sigma c} h^{\alpha\beta} \notag \\
    &-4\gamma_{3}\Zh^{\rho}_{a}\Zh^{\sigma}_{b}\Zs^{a}\Zs^{b}h^{\alpha\beta}-2\gamma_{2} \Zh^{\beta}_{c}\Zh^{\sigma c} h^{\alpha\rho}+2(\gamma_{4}-\omega_{5})\Zs^{2} \Zh^{\beta}_{c}\Zh^{\sigma c} h^{\alpha\rho} \notag \\
    & -2\omega_{3}\Zh^{\lambda}_{a}\Zh_{\lambda b}\Zh^{\beta a}\Zh^{\sigma b}  h^{\alpha\rho} -2(\gamma_{4}-\omega_{3}-\omega_{5})\Zh^{\beta}_{a}\Zh^{\sigma}_{b}\Zs^{a}\Zs^{b}h^{\alpha\rho}-2\gamma_{2} \Zh^{\beta}_{c}\Zh^{\rho c} h^{\alpha\sigma}+2(\gamma_{4}-\omega_{5})\Zs^{2} \Zh^{\beta}_{c}\Zh^{\rho c} h^{\alpha\sigma} \notag \\
    & -2\omega_{3}\Zh^{\lambda}_{a}\Zh_{\lambda b} \Zh^{\beta a} \Zh^{\rho b} h^{\alpha\sigma} -2(\gamma_{4}-\omega_{3}-\omega_{5})\Zh^{\beta}_{a}\Zh^{\rho}_{b}\Zs^{a}\Zs^{b} h^{\alpha \sigma} -2\gamma_{2} \Zh^{\alpha}_{c}\Zh^{\sigma c} h^{\beta\rho} +2(\gamma_{4}-\omega_{5})\Zs^{2}\Zh^{\alpha}_{c}\Zh^{\sigma c} h^{\beta \rho} \notag \\
    & -2\omega_{3}\Zh^{\lambda}_{a}\Zh_{\lambda b}\Zh^{\alpha a}\Zh^{\sigma b} h^{\beta\rho}-2(\gamma_{4}-\omega_{3}-\omega_{5})\Zh^{\alpha}_{a}\Zh^{\sigma}_{b}\Zs^{a}\Zs^{b}h^{\beta\rho} +2\gamma_{2}\Zs^{2}h^{\alpha\sigma}h^{\beta\rho}  +2(\gamma_{4}-\omega_{5})\Zs^{2} \Zh^{\alpha}_{c}\Zh^{\rho c} h^{\beta\sigma}\notag \\
    & -2\omega_{3}\Zh^{\lambda}_{a}\Zh_{\lambda b} \Zh^{\alpha a}\Zh^{\rho b} h^{\beta\sigma}-2(\gamma_{4}-\omega_{3}-\omega_{5})\Zh^{\alpha}_{a}\Zh^{\rho}_{b}\Zs^{a}\Zs^{b}h^{\beta\sigma} -2\gamma_{2} \Zh^{\alpha}_{c}\Zh^{\rho c} h^{\beta\sigma} \notag \\
    &  +2\omega_{3}\Zh_{a}^{\lambda}\Zh_{\lambda b}\Zs^{a}\Zs^{b} h^{\alpha\sigma}h^{\beta\rho}  -2\omega_{3}\Zs^{4}h^{\alpha\sigma}h^{\beta\rho} +\frac{1}{2}f h^{\alpha\sigma}h^{\beta\rho} +2\gamma_{2}\Zs^{2} h^{\alpha\rho}h^{\beta\sigma} \notag \\
    & -2\omega_{3}\Zs^{4} h^{\alpha\rho}h^{\beta\sigma}+2\omega_{3}\Zh_{a}^{\lambda}\Zh_{\lambda b}\Zs^{a}\Zs^{b} h^{\alpha\rho}h^{\beta\sigma}+\frac{1}{2}f h^{\alpha\rho}h^{\beta\sigma} \notag \\
    &+2(\gamma_{3}-\omega_{2})\Zs^{2}\Zh_{c}^{\alpha}\Zh^{\beta c}h^{\rho\sigma}-4\gamma_{3}\Zh_{a}^{\alpha}\Zh_{b}^{\beta}\Zs^{a}\Zs^{b}h^{\rho\sigma} +4\gamma_{1}\Zs^{2} h^{\alpha\beta}h^{\rho\sigma} \notag \\
    &-4\omega_{1}\Zs^{2}h^{\alpha\beta}h^{\rho\sigma}+4\omega_{1}\Zh_{a}^{\lambda}\Zh_{\lambda b}\Zs^{a}\Zs^{b} h^{\alpha\beta}h^{\rho\sigma} +2f_{X}\Zh_{c}^{\rho}\Zh^{\sigma c}h^{\alpha\beta}+2f_{X}\Zh_{c}^{\alpha}\Zh^{\beta c} h^{\rho\sigma} \notag \\
    &-\omega_{6}\Zh_{a}^{\lambda}\Zh_{b}^{\alpha_{1}}\Zh^{\beta a}\Zh^{\rho b}\varepsilon_{\lambda\alpha_{1}\alpha_{2}\alpha_{3}}h^{\alpha\alpha_{2}}h^{\sigma\alpha_{3}} -\omega_{6}\Zh_{a}^{\lambda}\Zh_{b}^{\alpha_{1}}\Zh^{\beta a}\Zh^{\sigma b}\varepsilon_{\lambda\alpha_{1}\alpha_{2}\alpha_{3}}h^{\alpha\alpha_{2}}h^{\rho\alpha_{3}} \notag \\ &-\omega_{6}\Zh_{a}^{\lambda}\Zh_{b}^{\alpha_{1}}\Zh^{\alpha a}\Zh^{\sigma b}\varepsilon_{\lambda\alpha_{1}\alpha_{2}\alpha_{3}}h^{\beta\alpha_{2}}h^{\rho\alpha_{3}} -\omega_{6}\Zh_{a}^{\lambda}\Zh_{b}^{\alpha_{1}}\Zh^{\alpha a}\Zh^{\rho b}\varepsilon_{\lambda\alpha_{1}\alpha_{2}\alpha_{3}}h^{\beta\alpha_{2}}h^{\sigma\alpha_{3}} \notag \\
    &-\omega_{6}\Zh_{a}^{\beta}\Zh_{b}^{\lambda}\Zh^{\sigma b}\Zs^{a}\varepsilon_{\lambda\alpha_{1}\alpha_{2}\alpha_{3}}h^{\alpha\alpha_{2}}h^{\rho\alpha_{3}}n^{\alpha_{1}}+\omega_{6}\Zh_{a}^{\beta}\Zh^{\lambda a}\Zh_{b}^{\sigma}\Zs^{b}\varepsilon_{\lambda\alpha_{1}\alpha_{2}\alpha_{3}}h^{\alpha\alpha_{2}}h^{\rho\alpha_{3}}n^{\alpha_{1}} \notag \\
    &-\omega_{6}\Zh_{a}^{\alpha}\Zh_{b}^{\lambda}\Zh^{\sigma b}\Zs^{a}\varepsilon_{\lambda\alpha_{1}\alpha_{2}\alpha_{3}}h^{\beta\alpha_{2}}h^{\rho\alpha_{3}}n^{\alpha_{1}}+\omega_{6}\Zh_{a}^{\alpha}\Zh^{\lambda a}\Zh_{b}^{\sigma}\Zs^{b}\varepsilon_{\lambda\alpha_{1}\alpha_{2}\alpha_{3}}h^{\beta\alpha_{2}}h^{\rho\alpha_{3}}n^{\alpha_{1}} \notag \\
    &-\omega_{6}\Zh_{a}^{\beta}\Zh_{b}^{\lambda}\Zh^{\rho b}\Zs^{a}\varepsilon_{\lambda\alpha_{1}\alpha_{2}\alpha_{3}}h^{\alpha\alpha_{2}}h^{\sigma\alpha_{3}}n^{\alpha_{1}}+\omega_{6}\Zh_{a}^{\beta}\Zh^{\lambda a}\Zh_{b}^{\rho}\Zs^{b}\varepsilon_{\lambda\alpha_{1}\alpha_{2}\alpha_{3}}h^{\alpha\alpha_{2}}h^{\sigma\alpha_{3}}n^{\alpha_{1}} \notag \\
    &-\omega_{6}\Zh_{a}^{\alpha}\Zh_{b}^{\lambda}\Zh^{\rho b}\Zs^{a}\varepsilon_{\lambda\alpha_{1}\alpha_{2}\alpha_{3}}h^{\beta\alpha_{2}}h^{\sigma\alpha_{3}}n^{\alpha_{1}}+\omega_{6}\Zh_{a}^{\alpha}\Zh^{\lambda a}\Zh_{b}^{\rho}\Zs^{b}\varepsilon_{\lambda\alpha_{1}\alpha_{2}\alpha_{3}}h^{\beta\alpha_{2}}h^{\sigma\alpha_{3}}n^{\alpha_{1}} \notag \\
    &-\omega_{6}\Zh_{a}^{\lambda}\Zh_{b}^{\alpha_{1}}\Zh^{\sigma a}\Zs^{b}\varepsilon_{\lambda\alpha_{1}\alpha_{2}\alpha_{3}}h^{\alpha\rho}h^{\beta\alpha_{3}}n^{\alpha_{2}} -\omega_{6}\Zh_{a}^{\lambda}\Zh_{b}^{\alpha_{1}}\Zh^{\rho a}\Zs^{b}\varepsilon_{\lambda\alpha_{1}\alpha_{2}\alpha_{3}}h^{\alpha\sigma}h^{\beta\alpha_{3}}n^{\alpha_{2}} \notag \\
    &-\omega_{6}\Zh_{a}^{\lambda}\Zh_{b}^{\alpha_{1}}\Zh^{\sigma a}\Zs^{b}\varepsilon_{\lambda\alpha_{1}\alpha_{2}\alpha_{3}}h^{\alpha\alpha_{3}}h^{\beta\rho}n^{\alpha_{2}} -\omega_{6}\Zh_{a}^{\lambda}\Zh_{b}^{\alpha_{1}}\Zh^{\rho a}\Zs^{b}\varepsilon_{\lambda\alpha_{1}\alpha_{2}\alpha_{3}}h^{\alpha\alpha_{3}}h^{\beta\sigma}n^{\alpha_{2}} \notag \\
    &-\omega_{6}\Zh_{a}^{\lambda}\Zh_{b}^{\alpha_{1}}\Zh^{\beta a}\Zs^{b}\varepsilon_{\lambda\alpha_{1}\alpha_{2}\alpha_{3}}h^{\alpha\sigma}h^{\rho\alpha_{3}}n^{\alpha_{2}} - \omega_{6}\Zh_{a}^{\lambda}\Zh_{b}^{\alpha_{1}}\Zh^{\alpha a}\Zs^{b}\varepsilon_{\lambda\alpha_{1}\alpha_{2}\alpha_{3}}h^{\beta\sigma}h^{\rho\alpha_{3}}n^{\alpha_{2}} \notag \\
    &-\omega_{6}\Zh_{a}^{\lambda}\Zh_{b}^{\alpha_{1}}\Zh^{\beta a}\Zs^{b}\varepsilon_{\lambda\alpha_{1}\alpha_{2}\alpha_{3}}h^{\alpha\rho}h^{\sigma\alpha_{3}}n^{\alpha_{2}}-\omega_{6}\Zh_{a}^{\lambda}\Zh_{b}^{\alpha_{1}}\Zh^{\alpha a}\Zs^{b}\varepsilon_{\lambda\alpha_{1}\alpha_{2}\alpha_{3}}h^{\beta\rho}h^{\sigma\alpha_{3}}n^{\alpha_{2}} \,,
\end{align}
where $X$, in this case, is $Z^{\mu}_a Z_{\mu}^a$ since we are considering the decoupling limit.
\section{Coefficients of the kinetic Lagrangian in \eq{Lkins}}\label{aD}
The elements of the matrix $\tilde{\mathcal{M}}_{1}$ in \eq{M1tilde} are given by
\begin{subequations}\label{termsM1}
\begin{align}
    \tilde{\A}_{1}&\equiv 4\left(\gamma_{1}+\gamma_{2}-(\gamma_{3} +\gamma_4+\omega_1+\omega_2+\omega_3+\omega_4+\omega_5)\Zs^{2}+(\omega_{1}+\omega_3)(\hat{\tilde{Z}}_{1}^{\ 1})^{2} \right) \,, \\
    \tilde{\A}_{2}&\equiv 4(\omega_{1}+\omega_{3})\hat{\tilde{Z}}_{1}^{\ 1} \hat{\tilde{Z}}_{1}^{\ 2} \,, \\
    \tilde{\A}_{3}&\equiv 4\ (\omega_{1}+\omega_{3})\hat{\tilde{Z}}_{1}^{\ 1} \sqrt{\Zh^{2}-(\hat{\tilde{Z}}_{1}^{\ 1})^{2}-(\hat{\tilde{Z}}_{1}^{\ 2})^{2}} \,, \\
    \tilde{\A}_{4}&\equiv 4\left(\gamma_{1}+\gamma_{2}-(\gamma_{3}+\gamma_4)\Zs^{2}+(\omega_{1}+\omega_3)(\hat{\tilde{Z}}_{1}^{\ 2})^{2} \right) \,, \\
    \tilde{\A}_{5}&\equiv 4(\omega_{1}+\omega_{3})\hat{\tilde{Z}}_{1}^{\ 2}\sqrt{\Zh^{2}-(\hat{\tilde{Z}}_{1}^{\ 1})^{2}-(\hat{\tilde{Z}}_{1}^{\ 2})^{2}} \,, \\
    \tilde{\A}_{6}&\equiv 4\left(\gamma_{1}+\gamma_{2}-(\gamma_{3}+\gamma_4)\Zs^{2}+(\omega_{1}+\omega_3)\Zh^{2}-(\omega_{1}+\omega_3)(\hat{\tilde{Z}}_{1}^{\ 1})^{2}-(\omega_{1}+\omega_3)(\hat{\tilde{Z}}_{1}^{\ 2})^{2}\right) \,, \\
    \tilde{\C}_{1} &\equiv 2\left(2\gamma_{1}+(\gamma_{3}-\omega_2-2\omega_4)\Zh^{2}-2(\gamma_{3}+\gamma_4-\omega_1+\omega_5)(\hat{\tilde{Z}}_{1}^{\ 1})^{2}-(\gamma_{3}+2\omega_1+\omega_2)\Zs^{2}+f_{X} \right)\sqrt{\Zs^{2}} \,, \\
    \tilde{\C}_{2}&\equiv 2\left(2\gamma_{1}-(\gamma_{3}+2\omega_1+\omega_2)\Zs^{2}+2\omega_{1}(\hat{\tilde{Z}}_{1}^{\ 1})^{2}+f_{X} \right)\sqrt{2\Zs^{2}} \,, \\
   \tilde{\C}_{3}&\equiv -4(\gamma_{3}+\gamma_{4}-\omega_{1}+\omega_{5})\hat{\tilde{Z}}_{1}^{\ 1}\hat{\tilde{Z}}_{1}^{\ 2}\sqrt{\Zs^{2}} \,, \\
    \tilde{\C}_{4}&\equiv 4\omega_{1}\sqrt{2\Zs^{2}}\hat{\tilde{Z}}_{1}^{\ 1}\hat{\tilde{Z}}_{1}^{\ 2} \,, \\
    \tilde{\C}_{5}&\equiv -4(\gamma_{3}+\gamma_{4}-\omega_{1}+\omega_{5})\hat{\tilde{Z}}_{1}^{\ 1}\sqrt{\Zs^{2}\left(\Zh^{2}-(\hat{\tilde{Z}}_{1}^{\ 1})^{2}-(\hat{\tilde{Z}}_{1}^{\ 2})^{2}\right)} \,, \\
    \tilde{\C}_{6}&\equiv 4\omega_{1}\hat{\tilde{Z}}_{1}^{\ 1}\sqrt{2\Zs^{2}\left(\Zh^{2}-(\hat{\tilde{Z}}_{1}^{\ 1})^{2}-(\hat{\tilde{Z}}_{1}^{\ 2})^{2}\right)} \,, \\
    \tilde{\F}_{1}&\equiv 4\left[(\gamma_1+\gamma_2)\Zs^{2} -(\omega_{1}+\omega_{3})\Zs^{4}
    -2\gamma_2\Zh^{2}+f_{X}\Zh^{2}-(\gamma_{4}+2\omega_{3}+\omega_{4}+\omega_{5})\Zh^{4} \right. \notag \\
    & \left. +\left((\gamma_{3}+2\gamma_4-\omega_2-2\omega_5)\Zh^{2}-(2\gamma_{3}+2\gamma_4-\omega_1-3\omega_3-2\omega_5)(\hat{\tilde{Z}}_{1}^{\ 1})^{2} \right)\Zs^{2} \right] \,, \\
    \tilde{\F}_{2}&\equiv \sqrt{2}\left[2\left(2\gamma_{1}+(\gamma_{3}-\omega_2)\Zh^{2}-2(\gamma_{3}-\omega_1)(\hat{\tilde{Z}}_{1}^{\ 1})^{2}-2\omega_{1}\Zs^{2}\right)\Zs^{2}+2f_{X}\Zh^{2}-f\right] \,, \\
    \tilde{\F}_{3}& \equiv 4\left[2\gamma_{1}+\gamma_{2}+(2\omega_{1}+\omega_{3})\left((\hat{\tilde{Z}}_{1}^{\ 1})^{2}-\Zs^{2}\right)\right]\Zs^{2}-f \,,
\end{align}
where $X$ is as in Appendix \ref{aC}.

\end{subequations}

On the other hand, the two non-zero entries of $\tilde{\mathcal{M}}_{2}$ in \eq{M2tilde} are given by
\begin{subequations}\label{termsM2}
\begin{align}
    \tilde{\F}_{4}&\equiv 4\left[\gamma_{2} \Zs^{2} -\gamma_2 \Zh^{2} -\omega_{3}\Zs^{4} -\omega_{3}\Zh^{4}-(\gamma_{4}-2\omega_{3}-\omega_{5})\Zs^{2} (\hat{\tilde{Z}}_{1}^{\ 1})^{2}+(\gamma_{4}-\omega_{5})\Zs^{2} \Zh^{2} \right]+f \,, \\
    \tilde{\F}_{5}&\equiv 4\left[\gamma_{2}\Zs^{2}+\omega_{3} \Zs^{2} \left((\hat{\tilde{Z}}_{1}^{\ 1})^{2}-\Zs^{2}\right) \right]+f \,.
\end{align}
\end{subequations}

\bibliography{Bibli.bib} 

\end{document}